\documentstyle[12pt,a4]{article}
\newcommand\slv{v\kern-5pt\raise1pt\hbox{$\scriptstyle/$}\kern1pt}

\newcommand\be{\begin{equation}}
\newcommand\ee{\end{equation}}
\newcommand{\eps}{\varepsilon}
\def\bq{\begin{eqnarray}}
\def\eq{\end{eqnarray}}

\def\mercutio{M{\sc ercutio}}
\def\menlo{M{\sc enlo} P{\sc arc}}
\def\debr{D{\sc ebrecen}}
\def\eerad{{\sc eerad}2}
\begin{document}

\thispagestyle{empty}
\begin{flushright}
NIKHEF-99-002\\
SPhT-T99/004 \\
%(Version of~\today)\\
\end{flushright}
\vspace{0.5cm}
\begin{center}
{\Large \bf QCD Corrections to Four-jet Production
and Three-jet Structure
in $e^+$ $e^-$ annihilation}\\[.3cm]
\vspace{1.7cm}
$\mbox{{\sc \bf Stefan Weinzierl$^{1,a}$ and David A. Kosower$^{2,b}$}}$
\begin{center} \em 
$^1$ NIKHEF, P.O. Box 41882, NL - 1009 DB Amsterdam, The Netherlands \\
\vspace{4mm}
$^2$ Service de Physique Th\'eorique, Centre d'Etudes de Saclay,\\ 
F-91191 Gif-sur-Yvette Cedex, France
\end{center}\end{center}
\vspace{2cm}

% abstract ---------------------------------------
\begin{abstract}\noindent
{
We report on the general purpose numerical program \mercutio, which can
be used to calculate any infrared safe four-jet quantity in electron-positron
annihilation at next-to-leading order.
The program is based on the dipole formalism and uses a remapping of phase-space in order
to improve the efficiency of the Monte Carlo integration.
Numerical results are given for the four-jet fraction and the $D$-parameter.
These results are compared with already existing ones in the literature
and serve as a cross-check.
The program can also be used to investigate the internal structure of three-jet events at
NLO. 
We give results for previously uncalculated observables: the jet broadening variable
and the softest-jet explanarity.
}    
\end{abstract}

\vspace*{\fill}

% footnotes -------------------------------------
\noindent 
$^a${\small email address : stefanw@nikhef.nl}\\
$^b${\small email address : kosower@spht.saclay.cea.fr}

% main text ------------------------------------
\newpage

\section{Introduction}

Electron-positron annihilation provides a clean environment for studying
jet
physics, as strong interaction effects occur only in the final state.
We thus avoid the complications of parton distribution functions and their
associated uncertainties.  Perturbative QCD
corrections to jet variables can nonetheless be sizeable.  The
most recent addition to the list of known one-loop amplitudes are those
 for $e^+ e^- \rightarrow 4$ partons, which have been
calculated by two groups independently.  Our group obtained analytic
expressions for the amplitudes \cite{amp1} by calculating helicity
amplitudes using the cut technique.  The Durham group performed a
conventional calculation and provided computer code \cite{amp3}.  The
two sets of expressions have been shown to agree numerically.  Here
we report on the implementation of the one-loop amplitudes into a
numerical program \mercutio, which can be used to calculate any
infrared safe four-, three- or two-jet quantity at next-to-leading
order (and five-jet quantities at leading order).  It is similar to
\menlo~\cite{res1}, \debr~\cite{res3} and 
\eerad~\cite{res9}.
The main difference among the programs is the technique used for the
cancellation of real and virtual singularities.  M{\sc enlo} P{\sc arc} 
uses the subtraction method of ref.~\cite{ir3}, \debr\ the dipole
method~\cite{ir1}, while \eerad\ uses a hybrid of the phase-space
slicing~\cite{ir2a,ir2} and subtraction methods.
\mercutio\ also uses the
dipole formalism, but with a remapping of phase
space.  A naive implementation of the dipole
formalism will give large statistical errors when performing a Monte
Carlo integration over the real corrections with dipole
factors subtracted. In order to improve the efficiency of the Monte Carlo
integration we remap the phase space in order to make the integrand 
more flat.
\\ 
\\ 
With the resulting numerical program, we first compute
results for several observables studied previously in the literature.
 We present results for the four-jet fraction and the
$D$-parameter, which agree with the results from the other programs.
We also present results for quantities describing the internal
structure of three-jet events.  We give numerical results for the jet
broadening and the softest-jet explanarity.
\\ 
\\ 
The paper is
organized as follows: in section 2 we explain the implementation of the
dipole formalism and the remapping of phase space.  Section~3 deals
with technical details concerning various dimensional regularization
schemes.  Section 4 is devoted to phenomenology, and contains
our numerical results.  The conclusions are given in section 5.

\section{Cancellation of Infrared Divergences}

The NLO cross section receives contributions both from virtual corrections
and from real emission.  For $e^+ e^- \rightarrow 4$ jets the virtual
part consists of the interference terms between the tree amplitudes
and the one-loop amplitudes for $e^+ e^- \rightarrow q g g \bar{q}$
and $e^+ e^- \rightarrow q \bar{q} q' \bar{q}'$.  The real emission part
consists of the squared tree-level amplitudes with one additional
parton, namely $e^+ e^- \rightarrow q g g g \bar{q}$ and $e^+ e^-
\rightarrow q \bar{q} q' \bar{q}' g$, which were computed long ago \cite{amp5}.  
Both the virtual and the real emission
contributions contain infrared divergences, arising when one particle
becomes soft or two particles collinear. Only the sum of the virtual
corrections and the real emission part is finite.  Because the virtual
part is integrated over $n$-particle phase-space, whereas the real
emission part is integrated over the $(n+1)$-particle phase space,
it is not possible
to integrate the divergent contributions numerically and to cancel the
divergences after the (numerical) integration.  The cancellation of
infrared singularities has to be performed before any numerical
integration is done.  There are basically two approaches to overcome
this obstacle: phase-space slicing \cite{ir2} and the subtraction
method.  (Within the subtraction method there are two variants: the
subtraction method by Frixione, Kunszt and Signer \cite{ir3} and the
dipole formalism of Catani and Seymour \cite{ir1}.) As mentioned
above, \mercutio\ uses the dipole formalism in order to handle infrared
divergences.

Throughout this paper the conventional normalization of the color
matrices is used: $\mbox{Tr} T^a T^b = \frac{1}{2} \delta^{ab}$.

\subsection{The dipole formalism}

The dipole formalism \cite{ir1} is based on the subtraction method. 
The NLO cross section is written as
\begin{eqnarray}
\sigma^{NLO} & = & \int\limits_{n+1} d\sigma^R + \int\limits_n d\sigma^V \nonumber \\
& = & \int\limits_{n+1} \left( d\sigma^R - d\sigma^A \right) 
       + \int\limits_n \left( d\sigma^V + \int\limits_1 
d\sigma^A \right)\,,
\end{eqnarray}
where in the last line an approximation term $d\sigma^A$ has been
added and subtracted.  The approximation $d\sigma^A$ has to be a
proper approximation of $d\sigma^R$ such as to have the same pointwise
singular behaviour in $D$ dimensions as $d\sigma^R$ itself.  Further
$d\sigma^A$ has to be analytically integrable in $D$ dimensions over
the one-parton subspace leading to soft and collinear divergences.
According to Catani and Seymour the approximation term is written as
the $n$-parton Born term $d\sigma^B$ times the dipole factors,
\begin{eqnarray}
d \sigma^A & = & \sum\limits_{\mbox{dipoles}} d\sigma^B \otimes
d V_{\mbox{dipole}}\,.
\end{eqnarray}
The integration over the $(n+1)$-parton phase space yields
\begin{eqnarray}
\int\limits_{n+1} d\sigma^A & = & \sum\limits_{\mbox{dipoles}}
\int\limits_n d\sigma^B \otimes \int\limits_1 d V_{\mbox{dipole}}
= \int\limits_n \left[ d\sigma^B \otimes I \right]\,,
\end{eqnarray}
where the universal factor $I$ is defined by
\begin{eqnarray}
I & = & \sum\limits_{\mbox{dipoles}} \int\limits_1 d V_{\mbox{dipoles}}\,.
\end{eqnarray}
For the explicit formul\ae\ we refer to the paper of Catani and Seymour \cite{ir1}.
Here, we discuss the implementation of the dipole formalism when using 
helicity amplitudes.
Let $A^\mu$ denote the $n$-parton amplitude, where the polarization
vector $\eps_\mu$ of the emitter gluon has been amputated.
Using
\begin{eqnarray}
\left(\eps_+^\mu\right)^\ast \eps_+^\nu + \left(\eps_-^\mu\right)^\ast \eps_-^\nu
& = & - g^{\mu\nu} + \frac{\tilde{p}_{ij}^\mu q^\nu + \tilde{p}_{ij}^\nu q^\mu}{\tilde{p}_{ij} \cdot q} ,
\end{eqnarray}
where $q$ is an arbitrary reference momentum, we obtain
\begin{eqnarray}
\left(A^\mu\right)^\ast \left(-g_{\mu\nu}\right) A^\nu & = & A_+^\ast A_+ + A_-^\ast A_- 
\end{eqnarray}
where $A_\pm$ denotes the helicity amplitude where the emitter gluon has positive or negative helicity.
The dependence on the reference momentum $q$ drops out, since $\left(\tilde{p}_{ij}\right)_\mu A^\mu = 0$ (gauge invariance),
$q_\mu \eps^\mu = 0$ (property of the polarization vectors) and $\left(\tilde{p}_{ij}\right)^2 = 0 $ (the gluon is on mass-shell).
For the spin correlation we obtain
\begin{eqnarray}
\left(A_\mu\right)^\ast \left(\tilde{z}_i p_i^\mu - \tilde{z}_j^\mu p_j^\mu \right)
\left(\tilde{z}_i p_i^\nu - \tilde{z}_j^\nu p_j^\nu \right) A_\nu
& = & \left| E A_+ + E^\ast A_- \right|^2,
\end{eqnarray}
where
\begin{eqnarray}
E & = & \frac{\langle q + | \tilde{z}_i p\!\!\!/_i - \tilde{z}_j p\!\!\!/_j | \tilde{p}_{ij} + \rangle}{\sqrt{2} [q \tilde{p}_{ij} ]}
\end{eqnarray}
Using the fact that the spin correlation tensor is orthogonal to $\tilde{p}_{ij}$ one shows again that the dependence
on the reference momentum drops out.
As reference momentum one may choose $q = p_j$, in that case $E$ reduces to
\begin{eqnarray}
E & = & \frac{\tilde{z}_i [ p_j p_i ] \langle p_i \tilde{p}_{ij} \rangle}{\sqrt{2} [p_j \tilde{p}_{ij} ]}
\end{eqnarray}

\subsection{Improving the Dipole Formalism}
\label{monte6}
Within the dipole formalism we have to evaluate the terms
\begin{eqnarray}
\int\limits_n \left( d\sigma^V + d\sigma^B \otimes I \right) & \mbox{and} & \int\limits_{n+1} d\sigma^R - d\sigma^A.
\end{eqnarray}
The first term is most efficiently integrated by splitting it
into leading-color and subleading-color pieces. The
leading-color piece gives the numerically dominant contribution. The
leading-color one-loop amplitudes are relatively simple and can
therefore be integrated without further problems. The subleading-color
one-loop amplitudes are more complicated and therefore need more
computer time. On the other hand, they will yield contributions
which are numerically smaller.  We can thus evaluate them 
using fewer integrand
evaluations.  Although the second term $d\sigma^R - d\sigma^A$ 
usually gives only a modest numerical contribution, a naive Monte Carlo
integration will give large statistical errors. Due to the large
number of dipole factors involved, it is also very computationally 
intensive.  We first deal with the fact that we have to reproduce
logarithms numerically.  This is solved by a remapping of
phase-space. In a second stage we use color decomposition in order to
reduce the number of dipole factors that have to be evaluated.\\

\def\ymin{y_{\rm min}}
A simplified model for the term $d\sigma^R - d\sigma^A$ would be
\begin{eqnarray}
F &= &\int\limits_0^1 dx \left( \frac{f(x)}{x} - \frac{g(x)}{x} \right)
\end{eqnarray}
where $f(0) = g(0)$ is assumed.
The integral can be rewritten as
\begin{eqnarray}
F & = & \int\limits_0^{\ymin} dx \frac{f(x)-g(x)}{x} 
 + \int\limits_{\ln \ymin}^0 dy \left( f(e^y) - g(e^y) \right)\,,
\end{eqnarray}
where $\ymin$ is an artificial parameter separating a numerically
dangerous region from a stable region.  Using the Taylor expansion for
$f(x)-g(x)$, one sees that the first term gives a contribution of
order $O(\ymin)$.  In the second term the $1/x$ behaviour has been
absorbed into the integral measure by a change of variables $y = \ln x$,
and the integrand tends to be more flat. This should reduce the
statistical error in a Monte Carlo integration.\\

The precise remapping of phase space for the term 
$d\sigma^R - d\sigma^A$ is done as follows: Consider a set of products of
invariants, including all the invariants in which the (unsubtracted)
matrix elements has poles (and which may give rise to logarithms
therefore).  The relevant set of products of invariants for the
two-quark, three-gluon final state consists of the pairs
\begin{eqnarray}
{\cal S}_g & = & \left\{ s_{q1} s_{12}, s_{q1} s_{13} , s_{q2} s_{21}
, s_{q2} s_{23}, s_{q3} s_{31}, s_{q3} s_{32} \right. \nonumber \\ 
& & s_{23} s_{3\bar{q}}, s_{32} s_{2\bar{q}}, s_{13} s_{3\bar{q}}, s_{31}
s_{1\bar{q}}, s_{12} s_{2\bar{q}}, s_{21} s_{1\bar{q}} \nonumber \\ 
& & s_{12} s_{23}, s_{13} s_{32}, s_{21} s_{13} \nonumber \\ 
& & \left. s_{q1} s_{1\bar{q}}, s_{q2} s_{2\bar{q}}, s_{q2} s_{2\bar{q}}
\right\},
\end{eqnarray}
where we have labelled the gluons from 1 to 3, $q$ corresponds to a quark and
$\bar{q}$ to the anti-quark.
Singularities associated with the last three pairs appear 
only in the color-subleading part.
The corresponding set for the four-quark, one-gluon final state consists of the pairs
\begin{eqnarray}
{\cal S}_q & = & \left\{ s_{qg} s_{g\bar{q}}, s_{q'g} s_{g\bar{q}'}, s_{qg} s_{g\bar{q}'}, s_{q'g} s_{g\bar{q}} \right. \nonumber \\
& & s_{q\bar{q}} s_{\bar{q}q'}, s_{q\bar{q}'} s_{\bar{q}'q'}, s_{\bar{q}q} s_{q\bar{q}'}, s_{\bar{q}q'} s_{q'\bar{q}'} \nonumber \\ 
& & \left. s_{qg} s_{gq'}, s_{\bar{q}g} s_{g\bar{q}'} \right\}.
\end{eqnarray}
The second line takes care of the collinear singularities when two
quarks become collinear.  The phase space is then partitioned
according to
\begin{eqnarray}
\Phi_{n+1} & = & \sum\limits_{{\cal S}} \Theta(a,b,c) \Phi_{n+1}
\end{eqnarray}
where $\Theta(a,b,c) = 1$ if $s_{ab} s_{bc}$ is the smallest product
in the set ${\cal S}$, and $\Theta(a,b,c) = 0$ otherwise. The sum is
over all products in the set ${\cal S}$.  We may use the symmetries
under permutations in order to reduce the number of summands which
need to be evaluated.  For the two-quark, three-gluon case we have to
evaluate the terms with $\Theta(q,1,2)$ and $\Theta(2,3,\bar{q})$ with
weight $6$, as well as the terms with $\Theta(1,2,3)$ and
$\Theta(q,3,\bar{q})$ which are weighted by a factor of $3$.
%Since the last term corresponds only to a subleading-color singularity we 
% may reduce the 
%relevant set of possible products in a leading color analysis and ignore the
%term with $\Theta(q,3,\bar{q})$.
For the four quark, one gluon case we evaluate the terms with
$\Theta(q,g,\bar{q}')$,$\Theta(q,g,\bar{q})$, $\Theta(q,\bar{q},q')$ and
$\Theta(\bar{q},q,\bar{q}')$, which are weighted by a factor $2$,
%the term with $\Theta(q,\bar{q},q')$ weighted by a factor $4$ 
and the terms with
$\Theta(q,g,q')$ and $\Theta(\bar{q},g,\bar{q}')$ with unit weight.\\

\def\smin{s_{\rm min}}
Next, introduce a parameter $\smin$ which separates numerically stable
and unstable regions.  (It is analogous to the $\smin$ parameter in
the phase-space slicing method, but plays a somewhat different role
here.)  Using this parameter, define two regions as follows.
Suppose 
 $s_{as} s_{sb}$ is the smallest product in the set.  Then
the first region is defined as the region
where $s_{as} > \smin$ and $s_{sb} > \smin$.
  In this region we perform a remapping of the
phase space as explained in detail in the next subsection.  \\

The second region is the complement of the first: 
$s_{as} < \smin$ or $s_{sb} < \smin$.
In this region the integration is performed without any phase space remapping.\\
\\
In summary, the numerical computation of the term $d\sigma^R - d\sigma^A$
goes as follows,
\begin{eqnarray}
d\sigma^R - d\sigma^A & = & \sum\limits_{{\cal S}} \left( d\sigma^R - d\sigma^A \right) 
 \Theta(s_{as} - \smin) \Theta(s_{sb} - \smin) \Theta(a,s,b) \nonumber \\
& & + \sum\limits_{{\cal S}} \left( d\sigma^R - d\sigma^A \right) 
 \left( 1- \Theta(s_{as} - \smin) \Theta(s_{sb} - \smin) \right) \Theta(a,s,b) \,.
\nonumber \\
\end{eqnarray}
Note that this separation into two regions
 does not involve any approximations and is exact whatever value $\smin$ might take. 
The aim is of course to choose $\smin$ so as to reduce 
the statistical errors.
It turns out that choosing $\smin$ small enough 
makes the contributions of the second region negligible.
The contribution of the first region has a statistical error 
reduced by roughly a factor of ten compared to the
result without any remapping and the same number of integrand evaluations. Empirically 
\begin{eqnarray}
\eta & = & \frac{\smin/Q^2}{y_{\rm cut}} = 10^{-5}
\end{eqnarray}
is a good value, where $y_{\rm cut}$ is the jet-defining parameter.

\subsection{Remapping of Phase Space}
\label{monte5}
The remapping of phase space we describe here was originally intended
for use with the phase space slicing method, and was used in prior jet
programs of one of the authors.  
The idea is to generate an $(n+1)$-parton configuration that is `near'
the hard $n$-parton configuration corresponding to the leading-order
calculation.  We want to do this in a way that makes the $(n+1)$-parton
configuration approach the hard configuration in singular limits.

Suppose we have a set of product of
invariants ${\cal S}$ such that $s_{as} s_{sb}$ is the smallest
product in the set.  In this region we remap the phase space as
follows.  Let $k_a'$, $k_s$ and $k_b'$ be the corresponding momenta
such that $s_{as} = (k_a' + k_s)^2$, $s_{sb} = (k_b' + k_s)^2$ and
$s_{ab} = (k_a' + k_s + k_b')^2$. We want to relate this $(n+1)$
particle configuration to a nearby ``hard'' $n$-particle configuration
with $(k_a + k_b)^2 = (k_a' + k_s + k_b')^2$, where $k_a$ and $k_b$
are the corresponding ``hard'' momenta.  Using the factorization of
the phase space, we have
\begin{eqnarray}
d\Phi_{n+1} & = & d\Phi_{n-1} \frac{dK^2}{2 \pi} d\Phi_3(K,k_a',k_s,k_b').
\end{eqnarray}
The three-particle phase space is given
by
\begin{eqnarray}
d\Phi_3(K,k_a',k_s,k_b') & = & \frac{1}{32 (2 \pi)^5 s_{ab}} ds_{as} ds_{sb} d\Omega_b' d\phi_s \nonumber \\
& = & \frac{1}{4 (2 \pi)^3 s_{ab}} ds_{as} ds_{sb} d\phi_s d\Phi_2(K,k_a,k_b)
\end{eqnarray}
and therefore
\begin{eqnarray}
d\Phi_{n+1} & = & d\Phi_n \frac{ds_{as} ds_{sb} d\phi_s}{4 (2 \pi)^3 s_{ab} }.
\end{eqnarray}
The region of integration for $s_{as}$ and $s_{sb}$ is 
$s_{as} > \smin$, $s_{sb} > \smin$ (either
from the $\Theta$-functions of phase space slicing method, or else
from the $\Theta$-functions separating the two regions above)
and $s_{as} + s_{sb} < s_{ab}$ (Dalitz plot for massless particles).
In order to get a flat integrand
we want to absorb poles in $s_{as}$ and $s_{sb}$ into the measure. 
This is done
by changing the variables according to
\begin{eqnarray}
s_{as} = s_{ab} \left( \frac{\smin}{s_{ab}} \right)^{u_1}, & &
s_{sb} = s_{ab} \left( \frac{\smin}{s_{ab}} \right)^{u_2} 
\end{eqnarray}
where $0 \leq u_1, u_2 \leq 1$. Note that $u_1, u_2 > 0$ 
enforces $s_{as}, s_{sb} > \smin$.
Therefore this transformation of variables may only be applied to invariants $s_{ij}$ 
where the region $0 < s_{ij} < \smin$ is cut out.
The phase space measure becomes
\begin{eqnarray}
d\Phi_{n+1} & = & d\Phi_{n} \frac{1}{4 (2 \pi)^3} \frac{s_{as} s_{sb}}{s_{ab}} \ln^2\left(\frac{\smin}{s_{ab}}\right)
\Theta(s_{as} + s_{sb} < s_{ab} ) du_1 du_2 d\phi_s .
\end{eqnarray}
This suggests the following algorithm for generating a $(n+1)$-parton configuration:
\begin{itemize}
\item Take a ``hard'' $n$-parton configuration and pick out two momenta $k_a$ and $k_b$. Use three
uniformly distributed (`random') numbers $u_1,u_2,u_3$ and set
\begin{eqnarray}
s_{ab} & = & (k_a + k_b)^2 ,\nonumber \\
s_{as} & = & s_{ab} \left( \frac{\smin}{s_{ab}} \right)^{u_1} ,\nonumber \\
s_{sb} & = & s_{ab} \left( \frac{\smin}{s_{ab}} \right)^{u_2} ,\nonumber \\
\phi_s & = & 2 \pi u_3.
\end{eqnarray}
\item If $(s_{as} + s_{sb} ) > s_{ab} $, reject the event.
\item If not, solve for $k_a'$, $k_b'$ and $k_s$.
If $s_{as} < s_{sb}$ we want to have $k_b' \rightarrow k_b$ as $s_{as} \rightarrow 0$.
Define
\begin{eqnarray}
E_a = \frac{s_{ab} - s_{sb}}{2 \sqrt{s_{ab}}}, \hspace{1cm}
E_s = \frac{s_{as}+s_{sb}}{2 \sqrt{s_{ab}}} ,\hspace{1cm}
E_b = \frac{s_{ab} - s_{as}}{2 \sqrt{s_{ab}}}, 
\end{eqnarray}
\begin{eqnarray}
\theta_{ab}  =  \arccos \left( 1 -\frac{s_{ab} - s_{as} - s_{sb}}{2 E_a E_b} \right), & &
\theta_{sb}  =  \arccos \left( 1 - \frac{s_{sb}}{2 E_s E_b} \right) .
\end{eqnarray}
It is convenient to work in a coordinate system which is obtained by a Lorentz transformation to the
center of mass of $k_a+k_b$ and a rotation such that $k_b'$ is along the positive $z$-axis. In that
coordinate system
\begin{eqnarray}
p_a' & = & E_a ( 1, \sin \theta_{ab} \cos(\phi_s+\pi), \sin \theta_{ab} \sin(\phi_s+\pi), \cos \theta_{ab} ) ,\nonumber \\
p_s & = & E_s ( 1, \sin \theta_{sb} \cos \phi_s, \sin \theta_{sb} \sin \phi_s, \cos \theta_{sb} ) ,\nonumber \\
p_b' & = & E_b ( 1, 0, 0, 1) .
\end{eqnarray}
The momenta $p_a'$, $p_s$ and $p_b'$ are related to the momenta $k_a'$, $k_s$ and $k_b'$ by a sequence of
Lorentz transformations back to the original frame
\begin{eqnarray}
k_a' & = & \Lambda_{boost} \Lambda_{xy}(\phi) \Lambda_{xz}(\theta) p_a'
\end{eqnarray}
and analogous for the other two momenta. 
The explicit formul\ae\ for the Lorentz transformations are obtained as follows :\\
\\
Let $K = \sqrt{(k_a+k_b)^2}$ and denote by $p_b$ the coordinates of the hard momentum $k_b$ in the center of
mass system of $k_a+k_b$. $p_b$ is given by
\begin{eqnarray}
p_b & = & \left( 
\frac{Q^0}{K} k_b^0 - \frac{\vec{k}_b \cdot \vec{Q}}{K}, \vec{k}_b + \left( \frac{\vec{k}_b \cdot \vec{Q}}{K (Q^0+K)} 
  - \frac{k_b^0}{K} \right) \vec{Q}
\right)
\end{eqnarray}
with $Q = k_a + k_b$. The angles are then given by
\begin{eqnarray}
\theta & = & \arccos \left( 1 - \frac{p_b \cdot p_b'}{2 p_b^t p_b^{t'}} \right), \nonumber \\
\phi & = & \arctan\left( \frac{p_b^y}{p_b^x} \right)
\end{eqnarray}
The explicit form of the rotations is
\begin{eqnarray}
\Lambda_{xz}(\theta) & = & \left(
\begin{array}{cccc}
1 & 0 & 0 & 0 \\
0 & \cos \theta & 0 & \sin \theta \\
0 & 0 & 1 & 0 \\
0 & - \sin \theta & 0 & \cos \theta \\
\end{array}
\right), \nonumber \\
\Lambda_{xy} (\phi) & = & 
\left(
\begin{array}{cccc}
1 & 0 & 0 & 0 \\
0 & \cos \phi & - \sin \phi & 0 \\
0 & \sin \phi & \cos \phi & 0 \\
0 & 0 & 0 & 1 \\
\end{array}
\right).
\end{eqnarray}
The boost $k' = \Lambda_{boost} q $ is given by 
\begin{eqnarray}
k' & = & \left( 
\frac{Q^0}{K} q^0 + \frac{\vec{q} \cdot \vec{Q}}{K}, \vec{q}+ \left( \frac{\vec{q} \cdot \vec{Q}}{K (Q^0+K)} 
  + \frac{q^0}{K} \right) \vec{Q}
\right)
\end{eqnarray}
with $Q = k_a + k_b$ and $K = \sqrt{(k_a+k_b)^2}$.
\item If $s_{as} > s_{sb}$, exchange $a$ and $b$ in the formul\ae\ above.
\item The ``soft'' event has then the weight
\begin{eqnarray}
W_{n+1} & = & \frac{\pi}{2} \frac{1}{(2 \pi)^3} \frac{s_{as} s_{sb}}{s_{ab}} \ln^2 \left( \frac{\smin}{s_{ab}} \right) W_n
\end{eqnarray}
where $W_n$ is the weight of the original ``hard'' event.
\end{itemize}

\subsection{Color decomposition and soft or collinear limits}

We now turn our attention to the color decomposition of the
five-parton tree amplitudes.  Singling out the numerical dominant
contribution, e.g. the leading color part, allows us to evaluate this
part with fewer dipole factors as subtraction terms.  This approach is
similar to the one followed within the phase-space slicing method.  In
fact, we have also implemented phase-space slicing for the leading
color contributions and checked that both methods, the dipole
formalism and phase-space slicing, give the same numerical results. As
a side-result we will give the relevant formul\ae\ for the contributions
from unresolved phase-space needed for the phase-space slicing
method.  

The color decomposition of the tree-level amplitude for 
$e^+ e^- \rightarrow q g_1 g_2 g_3 \bar{q}$ is
\begin{eqnarray}
A_5(q,g_1,g_2,g_3,\bar{q}) & = & ( T^1 T^2 T^3 )_{q\bar{q}} A_{5}^{(1)}(q,1,2,3,\bar{q})
        + ( T^1 T^3 T^2 )_{q\bar{q}} A_{5}^{(2)}(q,1,3,2,\bar{q}) \nonumber \\
& &     + ( T^2 T^3 T^1 )_{q\bar{q}} A_{5}^{(3)}(q,2,3,1,\bar{q})
        + ( T^2 T^1 T^3 )_{q\bar{q}} A_{5}^{(4)}(q,2,1,3,\bar{q}) \nonumber \\
& &     + ( T^3 T^1 T^2 )_{q\bar{q}} A_{5}^{(5)}(q,3,1,2,\bar{q})
        + ( T^3 T^2 T^1 )_{q\bar{q}} A_{5}^{(6)}(q,3,2,1,\bar{q}).
\end{eqnarray}
The index $i$ of $A_{5}^{(i)}$ labels the permutation of the gluons.
Explicit formul\ae\ for the partial amplitudes are given in ref.~\cite{amp5}.

The tree-level amplitude for $e^+ e^- \rightarrow g q \bar{q} q' \bar{q}'$ can be written as
\begin{eqnarray}
A_5(g,q,\bar{q},q',\bar{q}') & = & \frac{1}{2} \delta_{14} T_{32} D_1 + \frac{1}{2} \delta_{32} T_{14} D_2 
- \frac{1}{2} \delta_{12} T_{34} D_3 - \frac{1}{2} \delta_{34} T_{12} D_4 
\end{eqnarray}
where the factor $1/2$ is due
to the conventional normalization of $T^a$ and
\begin{eqnarray}
D_1 & = & c_0(1) B_1(0;1,2;3,4) + c_0(3) B_2(0;3,4;1,2) \nonumber \\
& & + \delta_{flav} \frac{1}{N_C} 
       \left( c_0(1) B_3(0;1,4;3,2) + c_0(3) B_4(0;3,2;1,4) \right), \nonumber \\
D_2 & = & c_0(1) B_2(0;1,2;3,4) + c_0(3) B_1(0;3,4;1,2) \nonumber\\
& &+ \delta_{flav} \frac{1}{N_C} 
       \left( c_0(1) B_4(0;1,4;3,2) + c_0(3) B_3(0;3,2;1,4) \right) ,\nonumber \\
D_3 & = & c_0(1) \frac{1}{N_C} B_3(0;1,2;3,4) + c_0(3) \frac{1}{N_C} B_4(0;3,4;1,2) \nonumber \\
& &+ \delta_{flav} 
       \left( c_0(1) B_1(0;1,4;3,2) + c_0(3) B_2(0;3,2;1,4) \right) ,\nonumber \\
D_4 & = & c_0(1) \frac{1}{N_C} B_4(0;1,2;3,4) + c_0(3) \frac{1}{N_C} B_3(0;3,4;1,2) \nonumber \\
& &+ \delta_{flav} 
       \left( c_0(1) B_2(0;1,4;3,2) + c_0(3) B_1(0;3,2;1,4) \right) .
\end{eqnarray}
Explicit formul\ae\ for the partial amplitudes $B_i$ are again
 given in ref.~\cite{amp5}.   In
this equation,
$c_0(j)$ denotes a factor from the electro-weak coupling and depends explicitly on the
flavor of the quark $q_j$,
\begin{eqnarray}
c_0(j) & = & -Q^{q_j} + v^e v^{q_j} {\cal P}_Z(s_{56})\,, \nonumber \\
{\cal P}_Z(s) & = & \frac{s}{s-M_Z^2+ i \Gamma_Z M_Z}\,.
\end{eqnarray}
The electron -- positron pair can either annihilate into a photon or a $Z$-boson. The first term in the expression
for $c_0$ corresponds to an intermediate photon, whereas the last term corresponds to a $Z$-boson.
The left- and right-handed couplings of the $Z$-boson to fermions are given by
\begin{eqnarray}
v_L = \frac{I_3 - Q \sin^2 \theta_W}{\sin \theta_W \cos \theta_W}, & &
v_R = \frac{- Q \sin \theta_W}{\cos \theta_W}
\end{eqnarray}
where $Q$ and $I_3$ are the charge and the third component of the weak isospin of the fermion. For an electron
and up- or down-type quarks we have:
\begin{center}
$
\begin{array}{ccc}
& Q & I_3 \\
& & \\
\left( \begin{array}{c} u \\ d \\ \end{array} \right) &
\left( \begin{array}{c} 2/3 \\ -1/3 \\ \end{array} \right) &
\left( \begin{array}{c} 1/2 \\ -1/2 \\ \end{array} \right) \\
& & \\
e^- & -1 & -1/2 \\
\end{array}
$
\end{center}
Next, consider the soft and collinear limit of any of these partial amplitudes,
e.g. $A_{5}^{(i)}$ or $B_i$.
In the soft-gluon limit,
a partial amplitude factorizes,
\begin{eqnarray}
A_{n+1} \rightarrow g \cdot \mbox{Eik}_{ab}^{\lambda} \cdot A_n\,,
\end{eqnarray}
where the eikonal factors are
\begin{eqnarray}
\mbox{Eik}_{ab}^+ = \sqrt{2} \frac{\langle ab \rangle}{\langle as \rangle \langle sb \rangle},
& &
\mbox{Eik}_{ab}^- = - \sqrt{2} \frac{ [ ab ] }{ [ as ] [ sb ] }.
\end{eqnarray}
Squaring the amplitude we obtain terms like
\begin{eqnarray}
\lefteqn{\sum\limits_{\mbox{soft gluon helicity $\lambda$}} A_{n+1}^{(2) \ast} A_{n+1}^{(1)} + A_{n+1}^{(1) \ast} A_{n+1}^{(2)} } & & \nonumber \\
& = & g^2 \left( \mbox{Eik}^{(2)-} \mbox{Eik}^{(1)+} + \mbox{Eik}^{(2)+} \mbox{Eik}^{(1)-} \right)
\left( A_n^{(2)\ast} A_n^{(1)} + A_n^{(1)\ast} A_n^{(2)} \right). 
\end{eqnarray}
There are three different cases, depending how the soft gluon is inserted into the partial amplitude
$A_{n+1}^{(1)}$ and $A_{n+1}^{(2)}$.
\begin{enumerate}
\item Two common legs, e.g. $A_{n+1}^{(1)} = A_{n+1}^{(1)}(...,a,s,b,...)$ and
$A_{n+1}^{(2)} = A_{n+1}^{(2)}(...,a,s,b,...)$, 
\begin{eqnarray}
\label{two_common_legs}
\sum\limits_{\lambda} A_{n+1}^{(2) \ast} A_{n+1}^{(1)} + A_{n+1}^{(1) \ast} A_{n+1}^{(2)} &  = &
g^2 \cdot 2 \cdot 2 \frac{s_{ab}}{s_{as} s_{sb}} \left( A_n^{(2)\ast} A_n^{(1)} + A_n^{(1)\ast} A_n^{(2)} \right) \nonumber \\
\end{eqnarray}
\item One common leg, e.g. $A_{n+1}^{(1)} = A_{n+1}^{(1)}(...,a,s,b,...)$ and
$A_{n+1}^{(2)} = A_{n+1}^{(2)}(...,a,s,c,...)$,
\begin{eqnarray}
\label{one_common_leg}
\lefteqn{
\sum\limits_{\lambda} A_{n+1}^{(2) \ast} A_{n+1}^{(1)} + A_{n+1}^{(1) \ast} A_{n+1}^{(2)} = }  & & \nonumber \\
&  = &
g^2 \cdot 2 \left( \frac{\langle ab \rangle}{\langle as \rangle \langle sb \rangle} \frac{[ ca ]}{[ cs ] [ sa ]}
+ \frac{[ ab ]}{[ as ] [ sb ]} \frac{\langle ca \rangle}{\langle cs \rangle \langle sa \rangle} \right)
\left( A_n^{(2)\ast} A_n^{(1)} + A_n^{(1)\ast} A_n^{(2)} \right) \nonumber \\
& = & g^2 \cdot 2 \left( \frac{s_{ab}}{s_{as} s_{sb}} -\frac{s_{bc}}{s_{bs} s_{sc}} + \frac{s_{ac}}{s_{as} s_{sc}} \right) 
\left( A_n^{(2)\ast} A_n^{(1)} + A_n^{(1)\ast} A_n^{(2)} \right) 
\end{eqnarray}
\item No common legs, e.g. $A_{n+1}^{(1)} = A_{n+1}^{(1)}(...,a,s,b,...)$ and
$A_{n+1}^{(2)} = A_{n+1}^{(2)}(...,d,s,c,...)$,
\begin{eqnarray}
\label{no_common_leg}
\lefteqn{
\sum\limits_{\lambda} A_{n+1}^{(2) \ast} A_{n+1}^{(1)} + A_{n+1}^{(1) \ast} A_{n+1}^{(2)} = } & & \nonumber \\
&  = &
g^2 \cdot 2 \left( \frac{\langle ab \rangle}{\langle as \rangle \langle sb \rangle} \frac{[ cd ]}{[ cs ] [ sd ]}
+ \frac{[ ab ]}{[ as ] [ sb ]} \frac{\langle cd \rangle}{\langle cs \rangle \langle sd \rangle} \right)
\left( A_n^{(2)\ast} A_n^{(1)} + A_n^{(1)\ast} A_n^{(2)} \right) \nonumber \\
& = & g^2 \cdot 2 \left( \frac{s_{ac}}{s_{as} s_{sc}} -\frac{s_{ad}}{s_{as} s_{sd}} + \frac{s_{bd}}{s_{bs} s_{sd}}
- \frac{s_{bc}}{s_{bs} s_{sc}} \right) 
\left( A_n^{(2)\ast} A_n^{(1)} + A_n^{(1)\ast} A_n^{(2)} \right) \nonumber \\
\end{eqnarray}
\end{enumerate}
In the collinear limit tree-level partial amplitudes factorize like
\begin{eqnarray}
A_{n+1} \rightarrow g \sum\limits_{\lambda = +/-} \mbox{Split}_{-\lambda} (p_a^{\lambda_a},p_b^{\lambda_b}) A_n(...,P^\lambda,...)
\end{eqnarray}
where $P = p_a + p_b$, $p_a = z P$ and $p_b = (1-z) P$. $\lambda$, $\lambda_a$ and $\lambda_b$ denote the
corresponding helicities.
The splitting functions are \cite{coll}:
\begin{eqnarray}
\mbox{Split}_{g^+}(g^+,g^+) = 0 & & \mbox{Split}_{g^-}(g^-,g^-) = 0 \nonumber \\
\mbox{Split}_{g^+}(g^+,g^-) = \sqrt{2} \frac{(1-z)^{\frac{3}{2}}}{\sqrt{z} \langle a b \rangle } & & 
 \mbox{Split}_{g^-}(g^-,g^+) = - \sqrt{2} \frac{(1-z)^{\frac{3}{2}}}{\sqrt{z} [ a b]} \nonumber \\
\mbox{Split}_{g^+}(g^-,g^+) = \sqrt{2} \frac{z^{\frac{3}{2}}}{\sqrt{(1-z)} \langle a b \rangle } & & 
 \mbox{Split}_{g^-}(g^+,g^-) = - \sqrt{2} \frac{z^{\frac{3}{2}}}{\sqrt{(1-z)} [ a b]} \nonumber \\
\mbox{Split}_{g^+}(g^-,g^-) = - \sqrt{2} \frac{1}{\sqrt{z(1-z)} [ a b ] } & & 
 \mbox{Split}_{g^-}(g^+,g^+) = \sqrt{2} \frac{1}{\sqrt{z(1-z)} \langle a b \rangle} \nonumber \\
& & \nonumber \\
\mbox{Split}_{q^-}(q^+,g^+) = \sqrt{2} \frac{1}{\sqrt{1-z} \langle a b \rangle} & & 
 \mbox{Split}_{q^+}(q^-,g^-) = -\sqrt{2} \frac{1}{\sqrt{1-z} [ a b ]} \nonumber \\
\mbox{Split}_{q^-}(q^+,g^-) = - \sqrt{2} \frac{z}{\sqrt{1-z} [ a b ]} & & 
 \mbox{Split}_{q^+}(q^-,g^+) = \sqrt{2} \frac{z}{\sqrt{1-z} \langle a b \rangle} \nonumber \\
\mbox{Split}_{\bar{q}^-}(g^+,\bar{q}^+) = \sqrt{2} \frac{1}{\sqrt{z} \langle a b \rangle} & & 
 \mbox{Split}_{\bar{q}^+}(g^-,\bar{q}^-) = -\sqrt{2} \frac{1}{\sqrt{z} [ a b ]} \nonumber \\
\mbox{Split}_{\bar{q}^-}(g^-,\bar{q}^+) = - \sqrt{2} \frac{1-z}{\sqrt{z} [ a b ]} & & 
 \mbox{Split}_{\bar{q}^+}(g^+,\bar{q}^-) = \sqrt{2} \frac{1-z}{\sqrt{z} \langle a b \rangle} \nonumber \\
& & \nonumber \\
\mbox{Split}_{g^+}(q^+, \bar{q}^-) = \sqrt{2} \frac{1-z}{\langle a b \rangle} & &
\mbox{Split}_{g^-}(q^-, \bar{q}^+) = -\sqrt{2} \frac{1-z}{[ a b ]} \nonumber \\
\mbox{Split}_{g^+}(q^-, \bar{q}^+) = - \sqrt{2} \frac{z}{\langle a b \rangle} & &
\mbox{Split}_{g^-}(q^+, \bar{q}^-) = \sqrt{2} \frac{z}{[ a b ]} 
\end{eqnarray}
We have checked numerically that all five-parton tree amplitudes in the program have the correct
collinear limits.

\subsection{Real Emission with Two Quarks and Three Gluons}

The matrix element squared of the tree-level amplitude for $e^+ e^- \rightarrow q g_1 g_2 g_3 \bar{q}$ can be written as
\begin{eqnarray} \label{Zqqggg}
| A_5(q,g_1,g_2,g_3,\bar{q}) |^2 & = & \vec{A}_5^\dagger C \vec{A}_5
\end{eqnarray}
where 
\begin{eqnarray}
\vec{A}_5 & = & \left( 
\begin{array}{c}
A_{5}^{(1)}(q,1,2,3,\bar{q}) \\ A_{5}^{(2)}(q,1,3,2,\bar{q}) \\ A_{5}^{(3)}(q,2,3,1,\bar{q}) \\
A_{5}^{(4)}(q,2,1,3,\bar{q}) \\ A_{5}^{(5)}(q,3,1,2,\bar{q}) \\ A_{5}^{(6)}(q,3,2,1,\bar{q})
\end{array} 
\right)
\end{eqnarray}
and the color matrix is given by
\begin{eqnarray}
C & = & \left(
\begin{array}{cccccc}
c_1^{(3)} & c_2^{(3)} & c_3^{(3)} & c_2^{(3)} & c_3^{(3)} & c_4^{(3)} \\
c_2^{(3)} & c_1^{(3)} & c_4^{(3)} & c_3^{(3)} & c_2^{(3)} & c_3^{(3)} \\
c_3^{(3)} & c_4^{(3)} & c_1^{(3)} & c_2^{(3)} & c_3^{(3)} & c_2^{(3)} \\
c_2^{(3)} & c_3^{(3)} & c_2^{(3)} & c_1^{(3)} & c_4^{(3)} & c_3^{(3)} \\
c_3^{(3)} & c_2^{(3)} & c_3^{(3)} & c_4^{(3)} & c_1^{(3)} & c_2^{(3)} \\
c_4^{(3)} & c_3^{(3)} & c_2^{(3)} & c_3^{(3)} & c_2^{(3)} & c_1^{(3)} \\
\end{array}
\right)
\end{eqnarray}
where the color factors are
\begin{eqnarray}
c_1^{(3)} & = & \left(\frac{1}{2}\right)^3 \frac{(N^2-1)^3}{N^2}, \nonumber \\
c_2^{(3)} & = & -\left(\frac{1}{2}\right)^3 \frac{(N^2-1)^2}{N^2}, \nonumber \\
c_3^{(3)} & = & \left(\frac{1}{2}\right)^3 \frac{N^2-1}{N^2}, \nonumber \\
c_4^{(3)} & = & \left(\frac{1}{2}\right)^3 \frac{(N^2-1) (N^2+1)}{N^2} = 2 c_3^{(3)} - c_2^{(3)}. 
\end{eqnarray}
If one gluon becomes soft or collinear, the matrix element squared reduces to the one with one gluon less
\begin{eqnarray} \label{lim_qqgg}
| A_5 |^2 & \rightarrow & \vec{A}_4^\dagger \left(
\begin{array}{cc}
c_1^{(2)} R_{diag}(q,1,2,\bar{q}) & c_2^{(2)} R_{off}(q,1,2,\bar{q}) \\
c_2^{(2)} R_{off}(q,1,2,\bar{q}) & c_1^{(2)} R_{diag}(q,2,1,\bar{q}) \\
\end{array}
\right)
\vec{A}_4 .
\end{eqnarray}
where the color factors are
\begin{eqnarray}
c_1^{(2)} & = & \left(\frac{1}{2} \right)^2 \frac{(N^2-1)^2}{N}, \nonumber \\
c_2^{(2)} & = & -\left(\frac{1}{2} \right)^2 \frac{N^2-1}{N} 
\end{eqnarray}
The finite $R$-factors, which give the contribution from unresolved phase-space
within the phase-space slicing method, are given by 
\begin{eqnarray}
\lefteqn{R_{diag}(q,1,2,\bar{q}) =} \nonumber \\
& = & \frac{g^2 c_\Gamma}{c_1^{(2)}} \left\{
2 c_1{(3)} R_2(q,1,2,\bar{q}) \right. \nonumber \\
& & -2 c_2^{(3)} \left( 2 R_2(q,1,2,\bar{q}) - R_1(q,1,\bar{q}) - R_1(q,2,\bar{q}) \right) \nonumber \\
& &+ 2 c_3^{(3)} \left( R_2(q,1,2,\bar{q}) - R_1(q,1,\bar{q}) - R_1(q,2,\bar{q})+ R_0(q,\bar{q})\right) \nonumber \\
& & \left. + 2 I_{g \rightarrow q \bar{q}}
\right\},  \nonumber \\
\lefteqn{R_{off}(q,1,2,\bar{q}) =} \nonumber \\
& = & \frac{g^2 c_\Gamma}{c_2^{(2)}} \left\{
-c_2^{(3)} \left( R_2(q,1,2,\bar{q}) + R_2(q,2,1,\bar{q}) - 2 R_1(q,1,\bar{q}) - 2 R_1(q,2,\bar{q})\right) \right. \nonumber \\
& & + 2 c_3^{(3)}  \left( R_2(q,1,2,\bar{q}) + R_2(q,2,1,\bar{q}) - R_1(q,1,\bar{q}) - R_1(q,2,\bar{q})\right) \nonumber \\
& & - c_4^{(3)}\left( R_2(q,1,2,\bar{q}) + R_2(q,2,1,\bar{q}) - 2 R_0(q,\bar{q}) \right) \nonumber \\
& & \left. + 2 I_{g \rightarrow q \bar{q}}
\right\}.
\end{eqnarray}
We have written the contribution from each color factor separately.
The last line corresponds to the contribution from the 
$A(g,q,\bar{q},q',\bar{q}')$ amplitudes, when one pair of quarks become collinear. 
We have used the notation
\begin{eqnarray}
R_0(q,\bar{q}) & = & S(q,\bar{q}) - \frac{4}{N_c} \frac{C_A}{2 C_F} 2 
I_{q \rightarrow q g}\left(\frac{\smin}{s_{q\bar{q}}}\right), \\
R_1(q,1,\bar{q}) & = & S(q,1) +S(1,\bar{q}) - \frac{4}{N_c} \left[ \frac{C_A}{2 C_F}  
I_{q \rightarrow q g}\left(\frac{\smin}{s_{q1}}\right)
\right. \nonumber \\
& & \left. + I_{g \rightarrow g g}\left(\frac{\smin}{s_{q1}},\frac{\smin}{s_{1\bar{q}}}\right) + \frac{C_A}{2 C_F}  
I_{q \rightarrow q g}\left(\frac{\smin}{s_{1\bar{q}}}\right) \right], \\
R_2(q,1,2,\bar{q}) & = & S(q,1) +S(1,2) +S(2,\bar{q}) - \frac{4}{N_c} \left[ \frac{C_A}{2 C_F}  
I_{q \rightarrow q g}\left(\frac{\smin}{s_{q1}}\right)
 \right. \nonumber \\
& & \left. + I_{g \rightarrow g g}\left(\frac{\smin}{s_{q1}},\frac{\smin}{s_{12}}\right)+ I_{g \rightarrow g g}\left(\frac{\smin}{s_{12}},\frac{\smin}{s_{2\bar{q}}}\right) \right. \nonumber \\
& & \left. + \frac{C_A}{2 C_F}  
I_{q \rightarrow q g}\left(\frac{\smin}{s_{2\bar{q}}}\right) \right] , \\
S(a,b) & = & \ln^2 \left( \frac{\mu^2}{\smin} \right) + \ln^2\left( \frac{s_{ab}}{\smin} \right)
+ 2 \ln \left( \frac{\mu^2}{\smin} \right) \ln \left( \frac{s_{ab}}{\smin} \right) ,\\
I_{a \rightarrow b c}(z_1,z_2) & = & I_{a \rightarrow b c}^1(z_1,z_2) +I_{a \rightarrow b c}^0(z_1,z_2)
\ln \left( \frac{\mu^2}{\smin} \right) .
\end{eqnarray}
$I^0_{a \rightarrow bc}$ and $I^1_{a \rightarrow bc}$ are the terms of order $O(\eps^0)$ and $O(\eps^1)$
obtained from the $D$-dimensional splitting functions integrated over unresolved phase space. 
In the conventional scheme they are given by
\begin{eqnarray}
I_{g \rightarrow g g} & = & \frac{C_A}{2} \left( - \ln z_1 - \ln z_2 - \frac{11}{6} +
 \left( \frac{1}{2} \ln^2 z_1 + \frac{1}{2} \ln^2 z_2 - \frac{67}{18} + \frac{\pi^2}{3} \right) \varepsilon 
+ O(\varepsilon^2) \right), \nonumber \\
I_{q \rightarrow q g} & = & C_F \left( - \ln z_2 - \frac{3}{4} + \left(
\frac{1}{2} \ln^2 z_2 -\frac{7}{4} + \frac{\pi^2}{6} \right) \varepsilon + O(\varepsilon^2) \right), \\
I_{g \rightarrow q \bar{q} } & = & T_R N_f \left( \frac{1}{3} + \frac{5}{9} \varepsilon + O(\varepsilon^2) \right).
\end{eqnarray}

\subsection{Real Emission with Four Quarks and One Gluon}

The matrix element squared of the tree-level amplitude for $e^+ e^- \rightarrow g q \bar{q} q' \bar{q}'$ can be written as
\begin{eqnarray} \label{Zqqqqg}
|A_5(g,q,\bar{q},q',\bar{q}') |^2 &=  & \vec{D}^{\dagger} C \vec{D}.
\end{eqnarray}
where 
\begin{eqnarray}
\vec{D} & = & \left( 
\begin{array}{c}
D_1 \\ D_2 \\ D_3 \\ D_4
\end{array} 
\right)
\end{eqnarray}
Then the color matrix is given by
\begin{eqnarray}
C & = &
\left(
\begin{array}{cccc}
c_1^{(1)} & 0 & c_2^{(1)} & c_2^{(1)} \\
0 & c_1^{(1)} & c_2^{(1)} & c_2^{(1)} \\
c_2^{(1)} & c_2^{(1)} & c_1^{(1)} & 0 \\
c_2^{(1)} & c_2^{(1)} & 0 & c_1^{(1)} \\
\end{array}
\right)
\end{eqnarray}
where
\begin{eqnarray}
c_1^{(1)} = \frac{1}{8} N_C ( N_C^2 -1 ), & & c_2^{(1)} = - \frac{1}{8} ( N_C^2-1)
\end{eqnarray}
The color decomposition of the four quark amplitude can be written as
\begin{eqnarray}
A_4 & = & \frac{1}{2} \left( \delta_{12} \delta_{34} 
- \frac{1}{N_c} \delta_{14} \delta_{32} \right) \chi(1,2,3,4) \nonumber \\
& & - \frac{1}{2} \left( \delta_{14} \delta_{32} - \frac{1}{N_c} \delta_{12} \delta_{34} \right) \delta_{flav}
\chi(1,4,3,2)
\end{eqnarray}
where
\begin{eqnarray}
\chi(1,2,3,4) = A_4(1,2,3,4) + A_4(3,4,1,2).
\end{eqnarray}
The matrix element squared we may write as
\begin{eqnarray}
|A_4|^2 & = & \vec{\chi}^\dagger \left(
\begin{array}{cc}
c_1^{(0)} & c_2^{(0)} \\
c_2^{(0)} & c_1^{(0)} \\
\end{array}
\right)
\vec{\chi}
\end{eqnarray}
with 
\begin{eqnarray}
c_1^{(0)} = \frac{1}{4} \left( N_c^2 -1 \right), & & c_2^{(0)} = \frac{1}{4} \frac{N_c^2 -1}{N_c}
\end{eqnarray}
and
\begin{eqnarray}
\vec{\chi} & = & \left(
\begin{array}{c}
\chi(1,2,3,4) \\
\delta_{flav} \chi(1,4,3,2) \\
\end{array}
\right).
\end{eqnarray}
In the soft gluon limit, the functions $B_i$ behave like
\begin{eqnarray}
B_1(0;1,2;3,4) & \rightarrow & g \cdot \mbox{Eik}_{32} \cdot A(1,2;3,4) ,\nonumber \\
B_2(0;1,2;3,4) & \rightarrow & g \cdot \mbox{Eik}_{14} \cdot A(1,2;3,4) ,\nonumber \\
B_3(0;1,2;3,4) & \rightarrow & g \cdot \mbox{Eik}_{34} \cdot A(1,2;3,4) ,\nonumber \\
B_4(0;1,2;3,4) & \rightarrow & g \cdot \mbox{Eik}_{12} \cdot A(1,2;3,4) .
\end{eqnarray}
In the limit where one pair of quarks becomes collinear, the amplitudes factorize as
\begin{eqnarray}
B_1(0;1,2;3,4) & \rightarrow & g \sum\limits_{\lambda} \mbox{Split}_{P}(3,4) A_4(1,P,0,2) ,\nonumber \\
B_2(0;1,2;3,4) & \rightarrow & g \sum\limits_{\lambda} \mbox{Split}_{P}(3,4) A_4(1,0,P,2) ,\nonumber \\
B_3(0;1,2;3,4) & \rightarrow & 0 ,\nonumber \\
B_4(0;1,2;3,4) & \rightarrow & g \sum\limits_{\lambda} \mbox{Split}_{P}(3,4) \left( A_4(1,P,0,2) + A_4(1,0,P,2) \right) .\nonumber \\
\end{eqnarray}
One of the functions $B_i$ is redundant, since
\begin{eqnarray}
B_1 + B_2 - B_3 - B_4 = 0 .
\end{eqnarray}
The contribution from unresolved phase space is written as
\begin{eqnarray} \label{lim_qqqq}
\vec{\chi}^\dagger \left(
\begin{array}{cc}
c_1^{(0)} R_{diag}(1,2,3,4) & c_2^{(0)} R_{off}(1,2,3,4) \\
c_2^{(0)} R_{off}(1,2,3,4) & c_1^{(0)} R_{diag}(1,4,3,2)\\
\end{array}
\right)
\vec{\chi}
\end{eqnarray}
with
\begin{eqnarray}
\lefteqn{R_{diag}(1,2,3,4) =} \nonumber \\
& = & \frac{g^2 c_\Gamma}{c_1^{(0)}} \left\{
2 c_1^{(1)} \left[ 
R_0(1,4) + R_0(2,3) + \frac{1}{N_c^2} \left( R_0(1,2) + R_0(3,4) \right) \right] \right. \nonumber \\
& & \left. + 4 \frac{c_2^{(1)}}{N_c}  \left[ R_0(1,4) + R_0(2,3) + R_0(1,2) + R_0(3,4) - R_0(1,3) - R_0(2,4) 
\right] \right\} ,\nonumber \\
\lefteqn{R_{off}(1,2,3,4) =} \nonumber \\
& = & \frac{g^2 c_\Gamma}{c_1^{(0)}} \left\{
2 \frac{c_1^{(1)}}{N_c} \left[ R_0(1,4) + R_0(2,3) + R_0(1,2) + R_0(3,4) \right] \right. \nonumber \\
& & \left. + 2 c_2^{(2)} \left( 1 + \frac{1}{N_c^2} \right)  \left[ R_0(1,4) + R_0(2,3) + R_0(1,2) + R_0(3,4) - R_0(1,3) - R_0(2,4) \right] \right\}
. \nonumber \\
\end{eqnarray}

\subsection{Color Correlation}

The color correlation matrices may be obtained by two ways. The first approach is the one given by Catani and
Seymour. 
Within this approach one acts with the color charge operators on the $n$-parton amplitudes. 
For example
\begin{eqnarray}
\lefteqn{
\left< q,1,2,\bar{q} | T_q \cdot T_1 | q,1,2,\bar{q} \right> =  
 \left(
T^2_{\bar{q}j} T^{1'}_{jq'} A(q,1,2,\bar{q})^\ast,
T^{1'}_{\bar{q}j} T^{2}_{jq'} A(q,2,1,\bar{q})^\ast
\right) } \nonumber \\
& & \cdot T^a_{q'q} \cdot 2 \mbox{Tr} \left( T^{1'} T^a T^1 - T^a T^{1'} T^1 \right) \cdot
\left(
\begin{array}{c}
T^1_{qi} T^2_{i\bar{q}} A(q,1,2,\bar{q}) \\
T^2_{qi} T^1_{i\bar{q}} A(q,2,1,\bar{q}) \\
\end{array}
\right)
\end{eqnarray}
Here the color charge operator for the gluon has been written as
\begin{eqnarray}
i f^{cab} & = & 2 \mbox{Tr} \left( T^c T^a T^b - T^a T^c T^b \right)
\end{eqnarray}
On the other hand one may start from the color decomposition of the $(n+1)$ -parton matrix element in the form
of equations (\ref{Zqqggg}) and (\ref{Zqqqqg}). One then considers the soft and collinear limits using the partial
fraction decompositions (\ref{two_common_legs}) - (\ref{no_common_leg}). This procedure is identical to the one
followed in the phase space slicing approach. With the help of the identity
\begin{eqnarray}
\frac{s_{ab}}{s_{as} s_{sb}} & = & \frac{s_{ab}}{s_{as} (s_{as} + s_{sb})}
                                  +\frac{s_{ab}}{s_{sb} (s_{as} + s_{sb})}
\end{eqnarray}
the color correlation matrix can then be read off from equations (\ref{lim_qqgg}) and (\ref{lim_qqqq}).
This approach has the advantage that it makes the connection between each divergent term in the $(n+1)$ -parton
matrix element and the corresponding subtraction term transparent.\\
\\
The leading order matrix element $e^+ e^- \rightarrow q g_1 g_2 g_3 \bar{q}$ needs 27 dipole factors.
There are six terms where the quark is the emitter and a gluon the spectator and three terms where the role of the
emitter and spectator are exchanged. The color correlation matrices are invariant under the exchange of emitter
and spectator. 
The color correlation matrix for the case $T_q \cdot T_1$ is given by
\begin{eqnarray}
T_q \cdot T_1 & = &
\left(
\begin{array}{cc}
c^{(3)}_2 - c^{(3)}_1 & \frac{1}{2} \left( c^{(3)}_4 - c^{(3)}_2 \right) \\
\frac{1}{2} \left( c^{(3)}_4 - c^{(3)}_2 \right) & c^{(3)}_3 - c^{(3)}_2 \\
\end{array}
\right)
\end{eqnarray}
If the antiquark $\bar{q}$ is the spectator we obtain 
\begin{eqnarray}
T_q \cdot T_{\bar{q}} & = &
\left(
\begin{array}{cc}
-c^{(3)}_3 & - c^{(3)}_4 \\
-c^{(3)}_4 & - c^{(3)}_3 \\
\end{array}
\right)
\end{eqnarray}
Finally the color correlation matrix where both emitter and spectator are gluons is given by
\begin{eqnarray}
T_1 \cdot T_2 & = &
\left(
\begin{array}{cc}
-c^{(3)}_1 + 2 c^{(3)}_2 -c^{(3)}_3 & 0 \\
0 & -c^{(3)}_1 + 2 c^{(3)}_2 -c^{(3)}_3  \\
\end{array}
\right)
\end{eqnarray}
All other color correlation matrices can be obtained by a permutation of indices.\\
\\
The leading order matrix element $e^+ e^- \rightarrow g q \bar{q} q' \bar{q}'$ needs 12 dipole factors
associated with the splitting $g \rightarrow q \bar{q}$ and 12 dipole factors associated to the splitting
$q \rightarrow q g$ or $\bar{q} \rightarrow g \bar{q}$. The relevant color correlation matrices for the first
case where already given above. The color correlation matrices for the latter case are
\begin{eqnarray}
T_q \cdot T_{\bar{q}} & = &
\left(
\begin{array}{cc}
-\frac{c^{(1)}_1}{N_c^2} -2 \frac{c^{(1)}_2}{N_c} & - \frac{c^{(1)}_1}{N_c} - c^{(1)}_2 \left( 1 + \frac{1}{N_c^2} \right) \\
- \frac{c^{(1)}_1}{N_c} - c^{(1)}_2 \left( 1 + \frac{1}{N_c^2} \right) & -c^{(1)}_1-2 \frac{c^{(1)}_2}{N_c}\\
\end{array}
\right)
\end{eqnarray}
\begin{eqnarray}
T_q \cdot T_{\bar{q}'} & = &
\left(
\begin{array}{cc}
-c^{(1)}_1-2 \frac{c^{(1)}_2}{N_c} & - \frac{c^{(1)}_1}{N_c} - c^{(1)}_2 \left( 1 + \frac{1}{N_c^2} \right) \\
- \frac{c^{(1)}_1}{N_c} - c^{(1)}_2 \left( 1 + \frac{1}{N_c^2} \right) & -\frac{c^{(1)}_1}{N_c^2} -2 \frac{c^{(1)}_2}{N_c}\\
\end{array}
\right)
\end{eqnarray}
\begin{eqnarray}
T_q \cdot T_{q'} & = &
\left(
\begin{array}{cc}
2 \frac{c^{(1)}_2}{N_c} &  c^{(1)}_2 \left( 1 + \frac{1}{N_c^2} \right) \\
c^{(1)}_2 \left( 1 + \frac{1}{N_c^2} \right) & 2 \frac{c^{(1)}_2}{N_c}\\
\end{array}
\right)
\end{eqnarray}
We can now pick out partial amplitudes corresponding to one specific color factor.
This subset will in general require fewer dipole factors as subtraction terms.
In practise it is enough to single out the leading color contribution for the matrix element
$e^+ e^- \rightarrow q g_1 g_2 g_3 \bar{q}$. Using the symmetry under the permutation of the
gluons it is sufficient to evaluate one partial amplitude, which we may take to be
\begin{eqnarray}
A_5^{(1)}(q,1,2,3,\bar{q}),
\end{eqnarray}
weighted by a factor of six.
This partial amplitude needs only six dipole factors $V_{ij,k}$, where $(ij)$ denotes
the emitter and $k$ the spectator:
\begin{eqnarray}
V_{q1,2}, \; V_{12,q}, \; V_{12,3}, \; V_{23,1}, \; V_{23,\bar{q}}, \; V_{3\bar{q},2}.
\end{eqnarray}
The color correlation matrices simplify accordingly, for example
\begin{eqnarray}
T_q \cdot T_1 & = &
\left(
\begin{array}{cc}
- c^{(3)}_1 & 0 \\
0 & 0 \\
\end{array}
\right),
\end{eqnarray}
which requires only the evaluation of one partial amplitude of the four-parton
tree-level amplitude.
We may further restrict the set of products of invariants to
\begin{eqnarray}
{\cal S}_g & =  & \left\{ s_{q1} s_{12}, s_{12} s_{23} , s_{23} s_{3\bar{q}}  
 \right\},
\end{eqnarray}
since we can only have poles in $s_{q1}$, $s_{12}$, $s_{23}$ or $s_{3\bar{q}}$.
Singling out the leading color contributions for
$e^+ e^- \rightarrow q g_1 g_2 g_3 \bar{q}$ together with 
the remapping of phase
space discussed in section~\ref{monte5}
improves the efficiency of the Monte Carlo integration of the
term $d\sigma^R - d\sigma^A$ sufficiently so that most of the computer time
is spent on the evaluation of the virtual one-loop amplitudes.\\
\\
We finally remark that Z. Nagy and Z. Tr\'ocs\'anyi \cite{res3} have chosen a different
approach in order to improve the efficiency of the dipole formalism: 
They introduce an additional parameter $\alpha$ and constrain the phase space
over which the dipole factors are subtracted:
\begin{eqnarray}
d\sigma^A = \sum\limits_{\mbox{dipoles}} d\sigma^B \otimes dV_{\mbox{dipole}}
\Theta(y_{ij,k} < \alpha)
\end{eqnarray}
where the dimensionless variable $y_{ij,k}$ is given by
\begin{eqnarray}
y_{ij,k} & = & \frac{p_i p_j}{p_i p_j + p_j p_k + p_i p_k}
\end{eqnarray}
The three methods (remapping of phase space, color decomposition
and restriction of the phase space for the dipole factors) could be combined,
if there is any need for further improvement in the efficiency.

\section{Regularization Schemes and Splitting Functions}

Theoretical calculations of infrared-safe quantities in QCD should
lead to unambiguous results, independent of the chosen regularization
scheme. 
Terms sensitive to the precise definition of the
regularization scheme enter the calculation usually in the virtual
part through tensor loop integrals, and in the real emission part
through the splitting functions \cite{split0}, which enter the dipole factors (in
the dipole formalism) or the contribution from unresolved phase space
( within phase space slicing).  
Within dimensional regularization the
most commonly used schemes are the conventional dimensional
regularization scheme (CDR) \cite{CDR}, where all momenta and all polarization vectors are taken
to be in $D$ dimensions,
the 't Hooft-Veltman scheme (HV) \cite{HV}, where the momenta and the helicities of the unobserved particles are $D$ dimensional,
whereas the momenta and the helicities of the observed particles are 4 dimensional,
and the four-dimensional helicity scheme (FDH) \cite{FDH}, where all polarization vectors are kept in four dimensions, as well
as the momenta of the observed particles. Only the momenta of the unobserved particles are continued to $D$
dimensions. \\
\\
The procedure adopted in the numerical program is as follows.
The one-loop amplitudes, calculated in the four-dimensional helicity scheme,
were converted to the HV scheme, using the transition rules~\cite{int12},
\begin{eqnarray}
A^{one-loop}_{FDH} - A^{one-loop}_{HV} & = & c_\Gamma g^2 A^{tree} \left( 
\frac{N_c}{3} - \frac{n_q}{4 N_c} + \frac{n_q N_c}{12} \right)
\end{eqnarray}
where $n_q$ is the number of quarks.
The one-loop
amplitude for $e^+e^- \rightarrow q \bar{q} g g $ is converted into the 't Hooft-Veltman scheme by
\begin{eqnarray}
A^{one-loop}_{HV} & = & A^{one-loop}_{FDH} - c_\Gamma \frac{1}{2} g^2 N_C \left(
1 - \frac{1}{N_C^2} \right)
A^{tree},
\end{eqnarray}
whereas the one-loop amplitude for $e^+ e^- \rightarrow q \bar{q} q' \bar{q}'$ is converted by
\begin{eqnarray}
A^{one-loop}_{HV} & = & A^{one-loop}_{FDH} - c_\Gamma g^2 N_C \left(
\frac{2}{3} - \frac{1}{N_C^2} \right)
A^{tree}.
\end{eqnarray}
The splitting functions entering the contribution from the integrals over the
dipole factors have then to be taken
in the 't Hooft-Veltman scheme as well. We take the formul\ae\ for the splitting functions in the HV-scheme
from Catani, Seymour and Tr\'ocs\'anyi 
%(which are identical to the ones in the CDR scheme) 
\cite{split1}.
They are given by
\begin{eqnarray}
P_{g \rightarrow g g} & = & 2 C_A \left( \frac{z}{1-z} + \frac{1-z}{z} + z (1-z) \right), \nonumber \\
P_{q \rightarrow q g} & = & 2 C_F \left( \frac{2 z}{1-z} + (1 -\varepsilon) ( 1- z) \right), \nonumber \\
P_{g \rightarrow q \bar{q}} & = & 2 T_R N_f \left( 1 - \frac{2}{1-\varepsilon} z (1-z)\right) 
\end{eqnarray}
A statistical
factor of $1/2!$ for two identical particles is included in the $g
\rightarrow g g$ case.  

\section{Phenomenology}

The numerical program \mercutio\ developed in this work
is a general purpose program for calculating any infrared safe four-, three- and
two-jet observable to next-to-leading order in $\alpha_s$
(and five-jet quantities at leading order). 
It is written in the language C++.
The only approximations which have been made are the neglect of the
light quark masses and terms which are suppressed by $1/m_{top}^4$
or higher powers of the top quark mass.
\mercutio\ calculates the quantities in fixed order in $\alpha_s$,
no resummation of terms of $\ln^2 y_{cut}$ or $\ln y_{cut}$ has (yet) been
implemented. Therefore our results are reliable only for 
values of $y_{cut}$ which are not too small.
The program calculates the jet quantities at the partonic level, and
no hadronization is done.\\
\\
The distribution for an infrared-safe variable $O$ at the center-of-mass energy $Q$
at next-to-leading order is given by two coefficients, $B_O$ and $C_O$, which
represent the leading and next-to-leading order perturbative
contributions:
\bq
\frac{1}{\sigma_{tot}} O \frac{d \sigma}{d O} & = & \left( \frac{\alpha_s}{2\pi} \right)^2 B_O
 + \left( \frac{\alpha_s}{2\pi} \right)^3 \left( C_O + 2 \beta_0 \ln \left( \frac{\mu^2}{Q^2} \right) B_O \right)
\eq
We normalize all distributions to the total hadronic cross section $\sigma_{tot}$
at $O(\alpha_s)$, given by
\begin{eqnarray}
\sigma_{tot} & = & \sigma_{2-jet}^{Born} \left( 1 + \frac{\alpha_s}{\pi} 
\right).
\end{eqnarray}
We further normalize each variable such that it takes values between 0 and 1. The average of the variable
is then defined as 
\begin{eqnarray}
\langle O \rangle & = & \frac{1}{\sigma_{tot}} \int\limits_0^1 O \frac{d\sigma}{d O} d O.
\end{eqnarray}
\\
As our nominal choice of input parameters we use $N_c=3$ colors and
$N_f =5$ massless quarks.
We take the electromagnetic coupling to be $\alpha(m_Z) = 1/127.9$ and the strong
coupling to be $\alpha_s(m_Z) = 0.118$. The numerical values of the $Z^0$-mass and width
are $m_Z = 91.187$ GeV and $\Gamma_Z = 2.490$ GeV. For the top
mass we take $m_t = 174$ GeV and for the weak mixing angle $\sin^2 \theta_W = 0.230$.
We take the center of mass energy to be $\sqrt{Q^2} = m_Z$ and we set
the renormalization scale equal to $\mu^2 = Q^2$.

\subsection{Four-Jet Fraction}

The four-jet fraction has been calculated by each group which has provided a numerical
NLO four-jet program and serves as a cross-check.
The four-jet fraction is defined as
\begin{eqnarray}
R_4 & = & \frac{\sigma_{4-jet}}{\sigma_{tot}}.
\end{eqnarray}
The values obtained for the four-jet fraction for different jet algorithms 
( JADE \cite{jet3}, DURHAM \cite{jet4} and GENEVA \cite{jet5} )
and varying $y_{cut}$ are given
in table 1, together with the corresponding values from the programs MENLO PARC by L. Dixon and A. Signer \cite{res1},
DEBRECEN by Z. Nagy and Z. Tr\'ocs\'anyi \cite{res3} and EERAD2 by J.M. Campbell, M.A. Cullen and E.W.N. Glover \cite{res9}.
\begin{table}
\label{jet_fraction}
\begin{center}
\begin{tabular}{|c|c|c|c|} \hline
Algorithm & $y_{cut}$ & \mercutio\ & MENLO PARC \\ \hline
        & 0.005 & $ (3.93 \pm 0.02 ) \cdot 10^{-1}$ & $ (3.79 \pm 0.08 ) \cdot 10^{-1} $ \\  
JADE-E0 & 0.01  & $ (1.93 \pm 0.01 ) \cdot 10^{-1} $ & $ (1.88 \pm 0.03 ) \cdot 10^{-1} $ \\  
        & 0.03  & $ (3.38 \pm 0.01 ) \cdot 10^{-2} $& $ (3.46 \pm 0.05 ) \cdot 10^{-2} $ \\ \hline  
 & $y_{cut}$ & DEBRECEN & EERAD2 \\ \hline
        & 0.005 & $ (3.88 \pm 0.07 ) \cdot 10^{-1} $ & $ (3.87 \pm 0.03 ) \cdot 10^{-1} $ \\  
        & 0.01  & $ (1.92 \pm 0.01 ) \cdot 10^{-1} $ & $ (1.93 \pm 0.01 ) \cdot 10^{-1} $ \\  
        & 0.03  & $ (3.37 \pm 0.01 ) \cdot 10^{-2} $ & $ (3.35 \pm 0.01 ) \cdot 10^{-2} $ \\ \hline  
& & & \\ \hline
Algorithm & $y_{cut}$ & \mercutio\ & MENLO PARC \\ \hline
       & 0.005 & $(1.06 \pm 0.01 ) \cdot 10^{-1}$ & $ (1.04 \pm 0.02 ) \cdot 10^{-1} $ \\  
DURHAM & 0.01  & $(4.72 \pm 0.01 ) \cdot 10^{-2}$ & $ (4.70 \pm 0.06 ) \cdot 10^{-2} $ \\  
       & 0.03  & $(6.96 \pm 0.03 ) \cdot 10^{-3}$ & $ (6.82 \pm 0.08 ) \cdot 10^{-3} $ \\ \hline  
 & $y_{cut}$ & DEBRECEN & EERAD2 \\ \hline
       & 0.005 & $ (1.05 \pm 0.01 ) \cdot 10^{-1} $ & $ (1.05 \pm 0.01 ) \cdot 10^{-1} $ \\  
       & 0.01  & $ (4.66 \pm 0.02 ) \cdot 10^{-2} $ & $ (4.65 \pm 0.02 ) \cdot 10^{-2} $ \\  
       & 0.03  & $ (6.87 \pm 0.04 ) \cdot 10^{-3} $ & $ (6.86 \pm 0.03 ) \cdot 10^{-3} $ \\ \hline  
& & & \\ \hline
Algorithm & $y_{cut}$ & \mercutio\ & MENLO PARC \\ \hline
       & 0.02 & $(2.67 \pm 0.05 ) \cdot 10^{-1}$ & $ (2.56 \pm 0.06 ) \cdot 10^{-1} $ \\  
GENEVA & 0.03 & $(1.79 \pm 0.03 ) \cdot 10^{-1}$ & $ (1.71 \pm 0.03 ) \cdot 10^{-1} $ \\  
       & 0.05 & $(8.53 \pm 0.07 ) \cdot 10^{-2}$ & $ (8.58 \pm 0.15 ) \cdot 10^{-2} $ \\ \hline  
 & $y_{cut}$ & DEBRECEN & EERAD2 \\ \hline
       & 0.02 & $ (2.63 \pm 0.06 ) \cdot 10^{-1} $ & $ (2.61 \pm 0.05 ) \cdot 10^{-1} $ \\  
       & 0.03 & $ (1.75 \pm 0.03 ) \cdot 10^{-1} $ & $ (1.72 \pm 0.03 ) \cdot 10^{-1} $ \\  
       & 0.05 & $ (8.37 \pm 0.12 ) \cdot 10^{-2} $ & $ (8.50 \pm 0.06 ) \cdot 10^{-2} $ \\ \hline  
\end{tabular}
\end{center}
\caption{The four-jet fraction as calculated by \mercutio, MENLO PARC, DEBRECEN
and EERAD2, for different jet algorithms and varying $y_{cut}$.}
\end{table}

\subsection{The D-Parameter}

Global event shape variables describe the topology of an event. They
may be calculated without reference to a jet defining algorithm. Like
jet algorithms they have to be infrared safe. An example of a global event shape variable is
the $D$-parameter \cite{pheno6} which is derived from the eigenvalues of the momentum tensor 
\begin{eqnarray}
\theta^{ij} & = & \frac{\sum\limits_a \frac{p_a^i p_a^j}{|\vec{p}_a|}}{\sum\limits_a |\vec{p}_a|}
\end{eqnarray}
where the sum runs over all final state particles and $p_a^i$ is the $i$-th component of the three-momentum
$\vec{p}_a$ of particle $a$ in the c.m. system. The tensor $\theta$ is normalized to have unit trace.
In terms of the eigenvalues of the $\theta$ tensor,
$\lambda_1$, $\lambda_2$, $\lambda_3$, with
$\lambda_1 + \lambda_2 + \lambda_3 = 1$, one defines
\begin{eqnarray}
D & = & 27 \lambda_1 \lambda_2 \lambda_3 =  27 \mbox{det} \theta^{ij}.
\end{eqnarray}
The range of values is $0 \leq D \leq 1$. The $D$-parameter measures
aplanarity, since one needs at least four final-state particles
to obtain a non-vanishing value.\\
\\
The values for the functions $B_D$ and $C_D$ are given in table 2.
Figure 1 shows the D-parameter distribution.
We remark that we have chosen to normalize the distribution to the total hadronic
cross-section at $O(\alpha_s)$.
The average of the shape variable is defined as
\begin{eqnarray}
\langle D \rangle & = & \frac{1}{\sigma_{tot}} \int\limits_0^1 D \frac{d\sigma}{d D} d D.
\end{eqnarray}
For the average we obtain
\begin{eqnarray}
\langle D \rangle & = & \left( \frac{\alpha_s}{2 \pi} \right)^2 \left( 5.82 \pm 0.01 \right) \cdot 10^{1}
                       +\left( \frac{\alpha_s}{2 \pi} \right)^3 \left( 2.43 \pm 0.06 \right) \cdot 10^{3}.
\end{eqnarray}
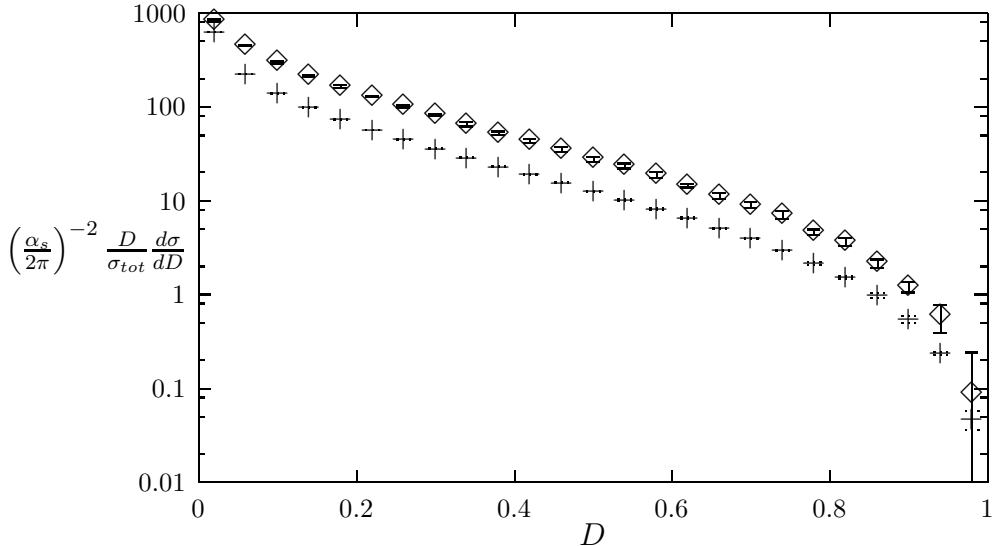
\begin{figure}
\begin{center}
%\epsfig{file=fig3.eps,height=10cm}
%\input fig_1_jet.tex
% GNUPLOT: LaTeX picture
\setlength{\unitlength}{0.240900pt}
\ifx\plotpoint\undefined\newsavebox{\plotpoint}\fi
\sbox{\plotpoint}{\rule[-0.200pt]{0.400pt}{0.400pt}}%
\begin{picture}(1500,900)(0,0)
\font\gnuplot=cmr10 at 10pt
\gnuplot
\sbox{\plotpoint}{\rule[-0.200pt]{0.400pt}{0.400pt}}%
\put(201.0,122.0){\rule[-0.200pt]{4.818pt}{0.400pt}}
\put(181,122){\makebox(0,0)[r]{0.01}}
\put(1420.0,122.0){\rule[-0.200pt]{4.818pt}{0.400pt}}
\put(201.0,166.0){\rule[-0.200pt]{2.409pt}{0.400pt}}
\put(1430.0,166.0){\rule[-0.200pt]{2.409pt}{0.400pt}}
\put(201.0,225.0){\rule[-0.200pt]{2.409pt}{0.400pt}}
\put(1430.0,225.0){\rule[-0.200pt]{2.409pt}{0.400pt}}
\put(201.0,255.0){\rule[-0.200pt]{2.409pt}{0.400pt}}
\put(1430.0,255.0){\rule[-0.200pt]{2.409pt}{0.400pt}}
\put(201.0,269.0){\rule[-0.200pt]{4.818pt}{0.400pt}}
\put(181,269){\makebox(0,0)[r]{0.1}}
\put(1420.0,269.0){\rule[-0.200pt]{4.818pt}{0.400pt}}
\put(201.0,314.0){\rule[-0.200pt]{2.409pt}{0.400pt}}
\put(1430.0,314.0){\rule[-0.200pt]{2.409pt}{0.400pt}}
\put(201.0,372.0){\rule[-0.200pt]{2.409pt}{0.400pt}}
\put(1430.0,372.0){\rule[-0.200pt]{2.409pt}{0.400pt}}
\put(201.0,403.0){\rule[-0.200pt]{2.409pt}{0.400pt}}
\put(1430.0,403.0){\rule[-0.200pt]{2.409pt}{0.400pt}}
\put(201.0,417.0){\rule[-0.200pt]{4.818pt}{0.400pt}}
\put(181,417){\makebox(0,0)[r]{1}}
\put(1420.0,417.0){\rule[-0.200pt]{4.818pt}{0.400pt}}
\put(201.0,461.0){\rule[-0.200pt]{2.409pt}{0.400pt}}
\put(1430.0,461.0){\rule[-0.200pt]{2.409pt}{0.400pt}}
\put(201.0,520.0){\rule[-0.200pt]{2.409pt}{0.400pt}}
\put(1430.0,520.0){\rule[-0.200pt]{2.409pt}{0.400pt}}
\put(201.0,550.0){\rule[-0.200pt]{2.409pt}{0.400pt}}
\put(1430.0,550.0){\rule[-0.200pt]{2.409pt}{0.400pt}}
\put(201.0,564.0){\rule[-0.200pt]{4.818pt}{0.400pt}}
\put(181,564){\makebox(0,0)[r]{10}}
\put(1420.0,564.0){\rule[-0.200pt]{4.818pt}{0.400pt}}
\put(201.0,609.0){\rule[-0.200pt]{2.409pt}{0.400pt}}
\put(1430.0,609.0){\rule[-0.200pt]{2.409pt}{0.400pt}}
\put(201.0,667.0){\rule[-0.200pt]{2.409pt}{0.400pt}}
\put(1430.0,667.0){\rule[-0.200pt]{2.409pt}{0.400pt}}
\put(201.0,697.0){\rule[-0.200pt]{2.409pt}{0.400pt}}
\put(1430.0,697.0){\rule[-0.200pt]{2.409pt}{0.400pt}}
\put(201.0,712.0){\rule[-0.200pt]{4.818pt}{0.400pt}}
\put(181,712){\makebox(0,0)[r]{100}}
\put(1420.0,712.0){\rule[-0.200pt]{4.818pt}{0.400pt}}
\put(201.0,756.0){\rule[-0.200pt]{2.409pt}{0.400pt}}
\put(1430.0,756.0){\rule[-0.200pt]{2.409pt}{0.400pt}}
\put(201.0,815.0){\rule[-0.200pt]{2.409pt}{0.400pt}}
\put(1430.0,815.0){\rule[-0.200pt]{2.409pt}{0.400pt}}
\put(201.0,845.0){\rule[-0.200pt]{2.409pt}{0.400pt}}
\put(1430.0,845.0){\rule[-0.200pt]{2.409pt}{0.400pt}}
\put(201.0,859.0){\rule[-0.200pt]{4.818pt}{0.400pt}}
\put(181,859){\makebox(0,0)[r]{1000}}
\put(1420.0,859.0){\rule[-0.200pt]{4.818pt}{0.400pt}}
\put(201.0,122.0){\rule[-0.200pt]{0.400pt}{4.818pt}}
\put(201,81){\makebox(0,0){0}}
\put(201.0,839.0){\rule[-0.200pt]{0.400pt}{4.818pt}}
\put(449.0,122.0){\rule[-0.200pt]{0.400pt}{4.818pt}}
\put(449,81){\makebox(0,0){0.2}}
\put(449.0,839.0){\rule[-0.200pt]{0.400pt}{4.818pt}}
\put(697.0,122.0){\rule[-0.200pt]{0.400pt}{4.818pt}}
\put(697,81){\makebox(0,0){0.4}}
\put(697.0,839.0){\rule[-0.200pt]{0.400pt}{4.818pt}}
\put(944.0,122.0){\rule[-0.200pt]{0.400pt}{4.818pt}}
\put(944,81){\makebox(0,0){0.6}}
\put(944.0,839.0){\rule[-0.200pt]{0.400pt}{4.818pt}}
\put(1192.0,122.0){\rule[-0.200pt]{0.400pt}{4.818pt}}
\put(1192,81){\makebox(0,0){0.8}}
\put(1192.0,839.0){\rule[-0.200pt]{0.400pt}{4.818pt}}
\put(1440.0,122.0){\rule[-0.200pt]{0.400pt}{4.818pt}}
\put(1440,81){\makebox(0,0){1}}
\put(1440.0,839.0){\rule[-0.200pt]{0.400pt}{4.818pt}}
\put(201.0,122.0){\rule[-0.200pt]{298.475pt}{0.400pt}}
\put(1440.0,122.0){\rule[-0.200pt]{0.400pt}{177.543pt}}
\put(201.0,859.0){\rule[-0.200pt]{298.475pt}{0.400pt}}
\put(41,490){\makebox(0,0){$\left( \frac{\alpha_s}{2 \pi}\right)^{-2} \frac{D}{\sigma_{tot}} \frac{d \sigma}{d D}$}}
\put(820,40){\makebox(0,0){$D$}}
\put(201.0,122.0){\rule[-0.200pt]{0.400pt}{177.543pt}}
\put(226.0,845.0){\rule[-0.200pt]{0.400pt}{0.964pt}}
\put(216.0,845.0){\rule[-0.200pt]{4.818pt}{0.400pt}}
\put(216.0,849.0){\rule[-0.200pt]{4.818pt}{0.400pt}}
\put(275.0,807.0){\rule[-0.200pt]{0.400pt}{0.482pt}}
\put(265.0,807.0){\rule[-0.200pt]{4.818pt}{0.400pt}}
\put(265.0,809.0){\rule[-0.200pt]{4.818pt}{0.400pt}}
\put(325.0,780.0){\rule[-0.200pt]{0.400pt}{0.723pt}}
\put(315.0,780.0){\rule[-0.200pt]{4.818pt}{0.400pt}}
\put(315.0,783.0){\rule[-0.200pt]{4.818pt}{0.400pt}}
\put(374.0,759.0){\rule[-0.200pt]{0.400pt}{0.723pt}}
\put(364.0,759.0){\rule[-0.200pt]{4.818pt}{0.400pt}}
\put(364.0,762.0){\rule[-0.200pt]{4.818pt}{0.400pt}}
\put(424.0,742.0){\rule[-0.200pt]{0.400pt}{0.964pt}}
\put(414.0,742.0){\rule[-0.200pt]{4.818pt}{0.400pt}}
\put(414.0,746.0){\rule[-0.200pt]{4.818pt}{0.400pt}}
\put(474.0,727.0){\usebox{\plotpoint}}
\put(464.0,727.0){\rule[-0.200pt]{4.818pt}{0.400pt}}
\put(464.0,728.0){\rule[-0.200pt]{4.818pt}{0.400pt}}
\put(523.0,711.0){\rule[-0.200pt]{0.400pt}{0.723pt}}
\put(513.0,711.0){\rule[-0.200pt]{4.818pt}{0.400pt}}
\put(513.0,714.0){\rule[-0.200pt]{4.818pt}{0.400pt}}
\put(573.0,697.0){\rule[-0.200pt]{0.400pt}{0.723pt}}
\put(563.0,697.0){\rule[-0.200pt]{4.818pt}{0.400pt}}
\put(563.0,700.0){\rule[-0.200pt]{4.818pt}{0.400pt}}
\put(622.0,681.0){\rule[-0.200pt]{0.400pt}{1.686pt}}
\put(612.0,681.0){\rule[-0.200pt]{4.818pt}{0.400pt}}
\put(612.0,688.0){\rule[-0.200pt]{4.818pt}{0.400pt}}
\put(672.0,668.0){\rule[-0.200pt]{0.400pt}{1.204pt}}
\put(662.0,668.0){\rule[-0.200pt]{4.818pt}{0.400pt}}
\put(662.0,673.0){\rule[-0.200pt]{4.818pt}{0.400pt}}
\put(721.0,655.0){\rule[-0.200pt]{0.400pt}{1.445pt}}
\put(711.0,655.0){\rule[-0.200pt]{4.818pt}{0.400pt}}
\put(711.0,661.0){\rule[-0.200pt]{4.818pt}{0.400pt}}
\put(771.0,641.0){\rule[-0.200pt]{0.400pt}{1.927pt}}
\put(761.0,641.0){\rule[-0.200pt]{4.818pt}{0.400pt}}
\put(761.0,649.0){\rule[-0.200pt]{4.818pt}{0.400pt}}
\put(821.0,625.0){\rule[-0.200pt]{0.400pt}{1.927pt}}
\put(811.0,625.0){\rule[-0.200pt]{4.818pt}{0.400pt}}
\put(811.0,633.0){\rule[-0.200pt]{4.818pt}{0.400pt}}
\put(870.0,615.0){\rule[-0.200pt]{0.400pt}{1.927pt}}
\put(860.0,615.0){\rule[-0.200pt]{4.818pt}{0.400pt}}
\put(860.0,623.0){\rule[-0.200pt]{4.818pt}{0.400pt}}
\put(920.0,600.0){\rule[-0.200pt]{0.400pt}{2.168pt}}
\put(910.0,600.0){\rule[-0.200pt]{4.818pt}{0.400pt}}
\put(910.0,609.0){\rule[-0.200pt]{4.818pt}{0.400pt}}
\put(969.0,585.0){\rule[-0.200pt]{0.400pt}{1.445pt}}
\put(959.0,585.0){\rule[-0.200pt]{4.818pt}{0.400pt}}
\put(959.0,591.0){\rule[-0.200pt]{4.818pt}{0.400pt}}
\put(1019.0,567.0){\rule[-0.200pt]{0.400pt}{2.409pt}}
\put(1009.0,567.0){\rule[-0.200pt]{4.818pt}{0.400pt}}
\put(1009.0,577.0){\rule[-0.200pt]{4.818pt}{0.400pt}}
\put(1068.0,553.0){\rule[-0.200pt]{0.400pt}{2.168pt}}
\put(1058.0,553.0){\rule[-0.200pt]{4.818pt}{0.400pt}}
\put(1058.0,562.0){\rule[-0.200pt]{4.818pt}{0.400pt}}
\put(1118.0,536.0){\rule[-0.200pt]{0.400pt}{2.891pt}}
\put(1108.0,536.0){\rule[-0.200pt]{4.818pt}{0.400pt}}
\put(1108.0,548.0){\rule[-0.200pt]{4.818pt}{0.400pt}}
\put(1167.0,511.0){\rule[-0.200pt]{0.400pt}{1.927pt}}
\put(1157.0,511.0){\rule[-0.200pt]{4.818pt}{0.400pt}}
\put(1157.0,519.0){\rule[-0.200pt]{4.818pt}{0.400pt}}
\put(1217.0,493.0){\rule[-0.200pt]{0.400pt}{3.132pt}}
\put(1207.0,493.0){\rule[-0.200pt]{4.818pt}{0.400pt}}
\put(1207.0,506.0){\rule[-0.200pt]{4.818pt}{0.400pt}}
\put(1267.0,459.0){\rule[-0.200pt]{0.400pt}{3.132pt}}
\put(1257.0,459.0){\rule[-0.200pt]{4.818pt}{0.400pt}}
\put(1257.0,472.0){\rule[-0.200pt]{4.818pt}{0.400pt}}
\put(1316.0,420.0){\rule[-0.200pt]{0.400pt}{4.095pt}}
\put(1306.0,420.0){\rule[-0.200pt]{4.818pt}{0.400pt}}
\put(1306.0,437.0){\rule[-0.200pt]{4.818pt}{0.400pt}}
\put(1366.0,357.0){\rule[-0.200pt]{0.400pt}{10.600pt}}
\put(1356.0,357.0){\rule[-0.200pt]{4.818pt}{0.400pt}}
\put(1356.0,401.0){\rule[-0.200pt]{4.818pt}{0.400pt}}
\put(1415.0,122.0){\rule[-0.200pt]{0.400pt}{49.144pt}}
\put(1405.0,122.0){\rule[-0.200pt]{4.818pt}{0.400pt}}
\put(226,847){\raisebox{-.8pt}{\makebox(0,0){$\Diamond$}}}
\put(275,808){\raisebox{-.8pt}{\makebox(0,0){$\Diamond$}}}
\put(325,782){\raisebox{-.8pt}{\makebox(0,0){$\Diamond$}}}
\put(374,761){\raisebox{-.8pt}{\makebox(0,0){$\Diamond$}}}
\put(424,744){\raisebox{-.8pt}{\makebox(0,0){$\Diamond$}}}
\put(474,727){\raisebox{-.8pt}{\makebox(0,0){$\Diamond$}}}
\put(523,713){\raisebox{-.8pt}{\makebox(0,0){$\Diamond$}}}
\put(573,699){\raisebox{-.8pt}{\makebox(0,0){$\Diamond$}}}
\put(622,684){\raisebox{-.8pt}{\makebox(0,0){$\Diamond$}}}
\put(672,670){\raisebox{-.8pt}{\makebox(0,0){$\Diamond$}}}
\put(721,658){\raisebox{-.8pt}{\makebox(0,0){$\Diamond$}}}
\put(771,645){\raisebox{-.8pt}{\makebox(0,0){$\Diamond$}}}
\put(821,630){\raisebox{-.8pt}{\makebox(0,0){$\Diamond$}}}
\put(870,619){\raisebox{-.8pt}{\makebox(0,0){$\Diamond$}}}
\put(920,605){\raisebox{-.8pt}{\makebox(0,0){$\Diamond$}}}
\put(969,588){\raisebox{-.8pt}{\makebox(0,0){$\Diamond$}}}
\put(1019,572){\raisebox{-.8pt}{\makebox(0,0){$\Diamond$}}}
\put(1068,557){\raisebox{-.8pt}{\makebox(0,0){$\Diamond$}}}
\put(1118,542){\raisebox{-.8pt}{\makebox(0,0){$\Diamond$}}}
\put(1167,515){\raisebox{-.8pt}{\makebox(0,0){$\Diamond$}}}
\put(1217,500){\raisebox{-.8pt}{\makebox(0,0){$\Diamond$}}}
\put(1267,466){\raisebox{-.8pt}{\makebox(0,0){$\Diamond$}}}
\put(1316,429){\raisebox{-.8pt}{\makebox(0,0){$\Diamond$}}}
\put(1366,383){\raisebox{-.8pt}{\makebox(0,0){$\Diamond$}}}
\put(1415,261){\raisebox{-.8pt}{\makebox(0,0){$\Diamond$}}}
\put(1405.0,326.0){\rule[-0.200pt]{4.818pt}{0.400pt}}
\put(226,830){\usebox{\plotpoint}}
\put(226,830){\usebox{\plotpoint}}
\put(216.00,830.00){\usebox{\plotpoint}}
\put(236,830){\usebox{\plotpoint}}
\put(216.00,830.00){\usebox{\plotpoint}}
\put(236,830){\usebox{\plotpoint}}
\put(275.00,763.00){\usebox{\plotpoint}}
\put(275,764){\usebox{\plotpoint}}
\put(265.00,763.00){\usebox{\plotpoint}}
\put(285,763){\usebox{\plotpoint}}
\put(265.00,764.00){\usebox{\plotpoint}}
\put(285,764){\usebox{\plotpoint}}
\put(325.00,733.00){\usebox{\plotpoint}}
\put(325,734){\usebox{\plotpoint}}
\put(315.00,733.00){\usebox{\plotpoint}}
\put(335,733){\usebox{\plotpoint}}
\put(315.00,734.00){\usebox{\plotpoint}}
\put(335,734){\usebox{\plotpoint}}
\put(374.00,711.00){\usebox{\plotpoint}}
\put(374,712){\usebox{\plotpoint}}
\put(364.00,711.00){\usebox{\plotpoint}}
\put(384,711){\usebox{\plotpoint}}
\put(364.00,712.00){\usebox{\plotpoint}}
\put(384,712){\usebox{\plotpoint}}
\put(424.00,692.00){\usebox{\plotpoint}}
\put(424,693){\usebox{\plotpoint}}
\put(414.00,692.00){\usebox{\plotpoint}}
\put(434,692){\usebox{\plotpoint}}
\put(414.00,693.00){\usebox{\plotpoint}}
\put(434,693){\usebox{\plotpoint}}
\put(474,676){\usebox{\plotpoint}}
\put(474,676){\usebox{\plotpoint}}
\put(464.00,676.00){\usebox{\plotpoint}}
\put(484,676){\usebox{\plotpoint}}
\put(464.00,676.00){\usebox{\plotpoint}}
\put(484,676){\usebox{\plotpoint}}
\put(523.00,660.00){\usebox{\plotpoint}}
\put(523,661){\usebox{\plotpoint}}
\put(513.00,660.00){\usebox{\plotpoint}}
\put(533,660){\usebox{\plotpoint}}
\put(513.00,661.00){\usebox{\plotpoint}}
\put(533,661){\usebox{\plotpoint}}
\put(573.00,646.00){\usebox{\plotpoint}}
\put(573,647){\usebox{\plotpoint}}
\put(563.00,646.00){\usebox{\plotpoint}}
\put(583,646){\usebox{\plotpoint}}
\put(563.00,647.00){\usebox{\plotpoint}}
\put(583,647){\usebox{\plotpoint}}
\put(622.00,632.00){\usebox{\plotpoint}}
\put(622,633){\usebox{\plotpoint}}
\put(612.00,632.00){\usebox{\plotpoint}}
\put(632,632){\usebox{\plotpoint}}
\put(612.00,633.00){\usebox{\plotpoint}}
\put(632,633){\usebox{\plotpoint}}
\put(672.00,618.00){\usebox{\plotpoint}}
\put(672,619){\usebox{\plotpoint}}
\put(662.00,618.00){\usebox{\plotpoint}}
\put(682,618){\usebox{\plotpoint}}
\put(662.00,619.00){\usebox{\plotpoint}}
\put(682,619){\usebox{\plotpoint}}
\put(721.00,605.00){\usebox{\plotpoint}}
\put(721,606){\usebox{\plotpoint}}
\put(711.00,605.00){\usebox{\plotpoint}}
\put(731,605){\usebox{\plotpoint}}
\put(711.00,606.00){\usebox{\plotpoint}}
\put(731,606){\usebox{\plotpoint}}
\put(771.00,592.00){\usebox{\plotpoint}}
\put(771,593){\usebox{\plotpoint}}
\put(761.00,592.00){\usebox{\plotpoint}}
\put(781,592){\usebox{\plotpoint}}
\put(761.00,593.00){\usebox{\plotpoint}}
\put(781,593){\usebox{\plotpoint}}
\put(821.00,578.00){\usebox{\plotpoint}}
\put(821,580){\usebox{\plotpoint}}
\put(811.00,578.00){\usebox{\plotpoint}}
\put(831,578){\usebox{\plotpoint}}
\put(811.00,580.00){\usebox{\plotpoint}}
\put(831,580){\usebox{\plotpoint}}
\put(870.00,564.00){\usebox{\plotpoint}}
\put(870,567){\usebox{\plotpoint}}
\put(860.00,564.00){\usebox{\plotpoint}}
\put(880,564){\usebox{\plotpoint}}
\put(860.00,567.00){\usebox{\plotpoint}}
\put(880,567){\usebox{\plotpoint}}
\put(920.00,550.00){\usebox{\plotpoint}}
\put(920,553){\usebox{\plotpoint}}
\put(910.00,550.00){\usebox{\plotpoint}}
\put(930,550){\usebox{\plotpoint}}
\put(910.00,553.00){\usebox{\plotpoint}}
\put(930,553){\usebox{\plotpoint}}
\put(969.00,536.00){\usebox{\plotpoint}}
\put(969,538){\usebox{\plotpoint}}
\put(959.00,536.00){\usebox{\plotpoint}}
\put(979,536){\usebox{\plotpoint}}
\put(959.00,538.00){\usebox{\plotpoint}}
\put(979,538){\usebox{\plotpoint}}
\put(1019.00,520.00){\usebox{\plotpoint}}
\put(1019,522){\usebox{\plotpoint}}
\put(1009.00,520.00){\usebox{\plotpoint}}
\put(1029,520){\usebox{\plotpoint}}
\put(1009.00,522.00){\usebox{\plotpoint}}
\put(1029,522){\usebox{\plotpoint}}
\put(1068.00,504.00){\usebox{\plotpoint}}
\put(1068,506){\usebox{\plotpoint}}
\put(1058.00,504.00){\usebox{\plotpoint}}
\put(1078,504){\usebox{\plotpoint}}
\put(1058.00,506.00){\usebox{\plotpoint}}
\put(1078,506){\usebox{\plotpoint}}
\put(1118.00,485.00){\usebox{\plotpoint}}
\put(1118,487){\usebox{\plotpoint}}
\put(1108.00,485.00){\usebox{\plotpoint}}
\put(1128,485){\usebox{\plotpoint}}
\put(1108.00,487.00){\usebox{\plotpoint}}
\put(1128,487){\usebox{\plotpoint}}
\put(1167.00,463.00){\usebox{\plotpoint}}
\put(1167,468){\usebox{\plotpoint}}
\put(1157.00,463.00){\usebox{\plotpoint}}
\put(1177,463){\usebox{\plotpoint}}
\put(1157.00,468.00){\usebox{\plotpoint}}
\put(1177,468){\usebox{\plotpoint}}
\put(1217.00,442.00){\usebox{\plotpoint}}
\put(1217,446){\usebox{\plotpoint}}
\put(1207.00,442.00){\usebox{\plotpoint}}
\put(1227,442){\usebox{\plotpoint}}
\put(1207.00,446.00){\usebox{\plotpoint}}
\put(1227,446){\usebox{\plotpoint}}
\put(1267.00,412.00){\usebox{\plotpoint}}
\put(1267,419){\usebox{\plotpoint}}
\put(1257.00,412.00){\usebox{\plotpoint}}
\put(1277,412){\usebox{\plotpoint}}
\put(1257.00,419.00){\usebox{\plotpoint}}
\put(1277,419){\usebox{\plotpoint}}
\put(1316.00,372.00){\usebox{\plotpoint}}
\put(1316,383){\usebox{\plotpoint}}
\put(1306.00,372.00){\usebox{\plotpoint}}
\put(1326,372){\usebox{\plotpoint}}
\put(1306.00,383.00){\usebox{\plotpoint}}
\put(1326,383){\usebox{\plotpoint}}
\put(1366.00,322.00){\usebox{\plotpoint}}
\put(1366,327){\usebox{\plotpoint}}
\put(1356.00,322.00){\usebox{\plotpoint}}
\put(1376,322){\usebox{\plotpoint}}
\put(1356.00,327.00){\usebox{\plotpoint}}
\put(1376,327){\usebox{\plotpoint}}
\multiput(1415,204)(0.000,20.756){2}{\usebox{\plotpoint}}
\put(1415,234){\usebox{\plotpoint}}
\put(1405.00,204.00){\usebox{\plotpoint}}
\put(1425,204){\usebox{\plotpoint}}
\put(1405.00,234.00){\usebox{\plotpoint}}
\put(1425,234){\usebox{\plotpoint}}
\put(226,830){\makebox(0,0){$+$}}
\put(275,763){\makebox(0,0){$+$}}
\put(325,734){\makebox(0,0){$+$}}
\put(374,711){\makebox(0,0){$+$}}
\put(424,693){\makebox(0,0){$+$}}
\put(474,676){\makebox(0,0){$+$}}
\put(523,661){\makebox(0,0){$+$}}
\put(573,646){\makebox(0,0){$+$}}
\put(622,632){\makebox(0,0){$+$}}
\put(672,618){\makebox(0,0){$+$}}
\put(721,606){\makebox(0,0){$+$}}
\put(771,592){\makebox(0,0){$+$}}
\put(821,579){\makebox(0,0){$+$}}
\put(870,566){\makebox(0,0){$+$}}
\put(920,551){\makebox(0,0){$+$}}
\put(969,537){\makebox(0,0){$+$}}
\put(1019,521){\makebox(0,0){$+$}}
\put(1068,505){\makebox(0,0){$+$}}
\put(1118,486){\makebox(0,0){$+$}}
\put(1167,466){\makebox(0,0){$+$}}
\put(1217,444){\makebox(0,0){$+$}}
\put(1267,416){\makebox(0,0){$+$}}
\put(1316,378){\makebox(0,0){$+$}}
\put(1366,325){\makebox(0,0){$+$}}
\put(1415,221){\makebox(0,0){$+$}}
\end{picture}
\caption{The D-parameter distribution at NLO (diamonds) and LO (crosses).}
\end{center}
\end{figure}
\begin{table}
\begin{center}
\begin{tabular}{|c|cc|} \hline
$D$ & $B_D$ & $C_D$ \\ \hline
0.02 & $ ( 6.36 \pm 0.04 ) \cdot 10^{2} $ & $ ( 1.05 \pm 0.13 ) \cdot 10^{4} $ \\
0.06 & $ ( 2.24 \pm 0.01 ) \cdot 10^{2} $ & $ ( 1.21 \pm 0.03 ) \cdot 10^{4} $ \\
0.10 & $ ( 1.41 \pm 0.01 ) \cdot 10^{2} $ & $ ( 8.37 \pm 0.26 ) \cdot 10^{3} $ \\
0.14 & $ ( 9.96 \pm 0.08 ) \cdot 10^{1} $ & $ ( 6.19 \pm 0.26 ) \cdot 10^{3} $ \\
0.18 & $ ( 7.43 \pm 0.04 ) \cdot 10^{1} $ & $ ( 4.84 \pm 0.25 ) \cdot 10^{3} $ \\
0.22 & $ ( 5.74 \pm 0.03 ) \cdot 10^{1} $ & $ ( 3.76 \pm 0.08 ) \cdot 10^{3} $ \\
0.26 & $ ( 4.51 \pm 0.04 ) \cdot 10^{1} $ & $ ( 3.00 \pm 0.13 ) \cdot 10^{3} $ \\
0.30 & $ ( 3.61 \pm 0.02 ) \cdot 10^{1} $ & $ ( 2.43 \pm 0.09 ) \cdot 10^{3} $ \\
0.34 & $ ( 2.90 \pm 0.02 ) \cdot 10^{1} $ & $ ( 1.93 \pm 0.19 ) \cdot 10^{3} $ \\
0.38 & $ ( 2.33 \pm 0.02 ) \cdot 10^{1} $ & $ ( 1.56 \pm 0.11 ) \cdot 10^{3} $ \\
0.42 & $ ( 1.91 \pm 0.03 ) \cdot 10^{1} $ & $ ( 1.29 \pm 0.11 ) \cdot 10^{3} $ \\
0.46 & $ ( 1.55 \pm 0.01 ) \cdot 10^{1} $ & $ ( 1.06 \pm 0.10 ) \cdot 10^{3} $ \\
0.50 & $ ( 1.26 \pm 0.02 ) \cdot 10^{1} $ & $ ( 8.08 \pm 0.93 ) \cdot 10^{2} $ \\
0.54 & $ ( 1.02 \pm 0.02 ) \cdot 10^{1} $ & $ ( 7.10 \pm 0.72 ) \cdot 10^{2} $ \\
0.58 & $ ( 8.20 \pm 0.16 ) \cdot 10^{0} $ & $ ( 5.64 \pm 0.64 ) \cdot 10^{2} $ \\
0.62 & $ ( 6.52 \pm 0.12 ) \cdot 10^{0} $ & $ ( 4.27 \pm 0.37 ) \cdot 10^{2} $ \\
0.66 & $ ( 5.09 \pm 0.08 ) \cdot 10^{0} $ & $ ( 3.32 \pm 0.46 ) \cdot 10^{2} $ \\
0.70 & $ ( 3.97 \pm 0.05 ) \cdot 10^{0} $ & $ ( 2.68 \pm 0.35 ) \cdot 10^{2} $ \\
0.74 & $ ( 2.94 \pm 0.06 ) \cdot 10^{0} $ & $ ( 2.19 \pm 0.35 ) \cdot 10^{2} $ \\
0.78 & $ ( 2.15 \pm 0.08 ) \cdot 10^{0} $ & $ ( 1.34 \pm 0.15 ) \cdot 10^{2} $ \\
0.82 & $ ( 1.52 \pm 0.05 ) \cdot 10^{0} $ & $ ( 1.14 \pm 0.20 ) \cdot 10^{2} $ \\
0.86 & $ ( 9.82 \pm 0.52 ) \cdot 10^{-1} $ & $ ( 6.28 \pm 1.15 ) \cdot 10^{1} $ \\
0.90 & $ ( 5.44 \pm 0.46 ) \cdot 10^{-1} $ & $ ( 3.55 \pm 0.79 ) \cdot 10^{1} $ \\
0.94 & $ ( 2.38 \pm 0.10 ) \cdot 10^{-1} $ & $ ( 1.87 \pm 1.03 ) \cdot 10^{1} $ \\
0.98 & $ ( 4.70 \pm 1.09 ) \cdot 10^{-2} $ & $ ( 2.15 \pm 8.31 ) \cdot 10^{0} $ \\ 
\hline
\end{tabular}
\end{center}
\caption{The Born level and next-to-leading order functions $B_D$ and $C_D$ for the $D$-parameter.}
\end{table}
These numbers agree with the ones given by Nagy and Tr\'ocs\'anyi as published 
in ref.~\cite{res3} 
after taking care of different normalizations ($\sigma_{tot}$ in our case and $\sigma^{Born}_{2-jet}$ in ref.~\cite{res3}). The $D$-parameter distribution
has also been calculated by Campbell, Cullen and Glover~\cite{res9}.

\subsection{The Jet Broadening Variable}

With the numerical program for $e^+ e^- \rightarrow \mbox{4 jets}$ one may also 
study the internal
structure of three-jets events. One example is the jet broadening variable
defined as \cite{aleph}
\begin{eqnarray}
B_{jet} & = & \frac{1}{n_{jets}} \sum\limits_{jets} \frac{\sum\limits_{a} |p_a^\perp |}{\sum\limits_a |\vec{p}_a|}
\end{eqnarray}
Here $p^\perp_a$ is the momentum of particle $a$ transverse to the jet axis of jet $J$, and the sum over $a$ extends over
all particles in the jet $J$. The jet broadening variable $B_{jet}$ considered here shall not be confused
with the narrow jet broadening $B_{min}$, which is defined differently. The later one
has been recently calculated to NLO by  Campbell, Cullen and Glover \cite{res9}.\\
\\
The jet broadening variable is calculated for three-jet events defined by the DURHAM algorithm and $y_{cut} = 0.1$. This choice is motivated by a recent analysis of the Aleph collaboration \cite{aleph}.
For the average we obtain
\begin{eqnarray}
\langle B_{jet} \rangle & = & \left( \frac{\alpha_s}{2 \pi} \right)^2 \left( 1.79 \pm 0.01 \right) \cdot 10^{1}
                             +\left( \frac{\alpha_s}{2 \pi} \right)^3 \left( 4.87 \pm 0.31 \right) \cdot 10^{2}.
\end{eqnarray}
\begin{figure}
\begin{center}
%\input fig_2_jet.tex
% GNUPLOT: LaTeX picture
\setlength{\unitlength}{0.240900pt}
\ifx\plotpoint\undefined\newsavebox{\plotpoint}\fi
\sbox{\plotpoint}{\rule[-0.200pt]{0.400pt}{0.400pt}}%
\begin{picture}(1500,900)(0,0)
\font\gnuplot=cmr10 at 10pt
\gnuplot
\sbox{\plotpoint}{\rule[-0.200pt]{0.400pt}{0.400pt}}%
\put(181.0,122.0){\rule[-0.200pt]{4.818pt}{0.400pt}}
\put(161,122){\makebox(0,0)[r]{-10}}
\put(1420.0,122.0){\rule[-0.200pt]{4.818pt}{0.400pt}}
\put(181.0,204.0){\rule[-0.200pt]{4.818pt}{0.400pt}}
\put(161,204){\makebox(0,0)[r]{0}}
\put(1420.0,204.0){\rule[-0.200pt]{4.818pt}{0.400pt}}
\put(181.0,286.0){\rule[-0.200pt]{4.818pt}{0.400pt}}
\put(161,286){\makebox(0,0)[r]{10}}
\put(1420.0,286.0){\rule[-0.200pt]{4.818pt}{0.400pt}}
\put(181.0,368.0){\rule[-0.200pt]{4.818pt}{0.400pt}}
\put(161,368){\makebox(0,0)[r]{20}}
\put(1420.0,368.0){\rule[-0.200pt]{4.818pt}{0.400pt}}
\put(181.0,450.0){\rule[-0.200pt]{4.818pt}{0.400pt}}
\put(161,450){\makebox(0,0)[r]{30}}
\put(1420.0,450.0){\rule[-0.200pt]{4.818pt}{0.400pt}}
\put(181.0,531.0){\rule[-0.200pt]{4.818pt}{0.400pt}}
\put(161,531){\makebox(0,0)[r]{40}}
\put(1420.0,531.0){\rule[-0.200pt]{4.818pt}{0.400pt}}
\put(181.0,613.0){\rule[-0.200pt]{4.818pt}{0.400pt}}
\put(161,613){\makebox(0,0)[r]{50}}
\put(1420.0,613.0){\rule[-0.200pt]{4.818pt}{0.400pt}}
\put(181.0,695.0){\rule[-0.200pt]{4.818pt}{0.400pt}}
\put(161,695){\makebox(0,0)[r]{60}}
\put(1420.0,695.0){\rule[-0.200pt]{4.818pt}{0.400pt}}
\put(181.0,777.0){\rule[-0.200pt]{4.818pt}{0.400pt}}
\put(161,777){\makebox(0,0)[r]{70}}
\put(1420.0,777.0){\rule[-0.200pt]{4.818pt}{0.400pt}}
\put(181.0,859.0){\rule[-0.200pt]{4.818pt}{0.400pt}}
\put(161,859){\makebox(0,0)[r]{80}}
\put(1420.0,859.0){\rule[-0.200pt]{4.818pt}{0.400pt}}
\put(181.0,122.0){\rule[-0.200pt]{0.400pt}{4.818pt}}
\put(181,81){\makebox(0,0){0}}
\put(181.0,839.0){\rule[-0.200pt]{0.400pt}{4.818pt}}
\put(433.0,122.0){\rule[-0.200pt]{0.400pt}{4.818pt}}
\put(433,81){\makebox(0,0){0.2}}
\put(433.0,839.0){\rule[-0.200pt]{0.400pt}{4.818pt}}
\put(685.0,122.0){\rule[-0.200pt]{0.400pt}{4.818pt}}
\put(685,81){\makebox(0,0){0.4}}
\put(685.0,839.0){\rule[-0.200pt]{0.400pt}{4.818pt}}
\put(936.0,122.0){\rule[-0.200pt]{0.400pt}{4.818pt}}
\put(936,81){\makebox(0,0){0.6}}
\put(936.0,839.0){\rule[-0.200pt]{0.400pt}{4.818pt}}
\put(1188.0,122.0){\rule[-0.200pt]{0.400pt}{4.818pt}}
\put(1188,81){\makebox(0,0){0.8}}
\put(1188.0,839.0){\rule[-0.200pt]{0.400pt}{4.818pt}}
\put(1440.0,122.0){\rule[-0.200pt]{0.400pt}{4.818pt}}
\put(1440,81){\makebox(0,0){1}}
\put(1440.0,839.0){\rule[-0.200pt]{0.400pt}{4.818pt}}
\put(181.0,122.0){\rule[-0.200pt]{303.293pt}{0.400pt}}
\put(1440.0,122.0){\rule[-0.200pt]{0.400pt}{177.543pt}}
\put(181.0,859.0){\rule[-0.200pt]{303.293pt}{0.400pt}}
\put(41,490){\makebox(0,0){\shortstack{$\left( \frac{\alpha_s}{2 \pi}\right)^{-2} \frac{1}{\sigma_{tot}} \;\;\;\;\;$ \\ $ \cdot \frac{B_{jet}}{n_{jets}} \frac{d \sigma}{d B_{jet}}\;\;\;\;\;$}}}
\put(810,40){\makebox(0,0){$B_{jet}$}}
\put(181.0,122.0){\rule[-0.200pt]{0.400pt}{177.543pt}}
\put(206.0,437.0){\rule[-0.200pt]{0.400pt}{49.625pt}}
\put(196.0,437.0){\rule[-0.200pt]{4.818pt}{0.400pt}}
\put(196.0,643.0){\rule[-0.200pt]{4.818pt}{0.400pt}}
\put(257.0,723.0){\rule[-0.200pt]{0.400pt}{20.717pt}}
\put(247.0,723.0){\rule[-0.200pt]{4.818pt}{0.400pt}}
\put(247.0,809.0){\rule[-0.200pt]{4.818pt}{0.400pt}}
\put(307.0,718.0){\rule[-0.200pt]{0.400pt}{14.936pt}}
\put(297.0,718.0){\rule[-0.200pt]{4.818pt}{0.400pt}}
\put(297.0,780.0){\rule[-0.200pt]{4.818pt}{0.400pt}}
\put(357.0,678.0){\rule[-0.200pt]{0.400pt}{11.804pt}}
\put(347.0,678.0){\rule[-0.200pt]{4.818pt}{0.400pt}}
\put(347.0,727.0){\rule[-0.200pt]{4.818pt}{0.400pt}}
\put(408.0,652.0){\rule[-0.200pt]{0.400pt}{10.359pt}}
\put(398.0,652.0){\rule[-0.200pt]{4.818pt}{0.400pt}}
\put(398.0,695.0){\rule[-0.200pt]{4.818pt}{0.400pt}}
\put(458.0,623.0){\rule[-0.200pt]{0.400pt}{10.840pt}}
\put(448.0,623.0){\rule[-0.200pt]{4.818pt}{0.400pt}}
\put(448.0,668.0){\rule[-0.200pt]{4.818pt}{0.400pt}}
\put(508.0,583.0){\rule[-0.200pt]{0.400pt}{9.154pt}}
\put(498.0,583.0){\rule[-0.200pt]{4.818pt}{0.400pt}}
\put(498.0,621.0){\rule[-0.200pt]{4.818pt}{0.400pt}}
\put(559.0,561.0){\rule[-0.200pt]{0.400pt}{5.059pt}}
\put(549.0,561.0){\rule[-0.200pt]{4.818pt}{0.400pt}}
\put(549.0,582.0){\rule[-0.200pt]{4.818pt}{0.400pt}}
\put(609.0,534.0){\rule[-0.200pt]{0.400pt}{5.300pt}}
\put(599.0,534.0){\rule[-0.200pt]{4.818pt}{0.400pt}}
\put(599.0,556.0){\rule[-0.200pt]{4.818pt}{0.400pt}}
\put(659.0,506.0){\rule[-0.200pt]{0.400pt}{3.854pt}}
\put(649.0,506.0){\rule[-0.200pt]{4.818pt}{0.400pt}}
\put(649.0,522.0){\rule[-0.200pt]{4.818pt}{0.400pt}}
\put(710.0,468.0){\rule[-0.200pt]{0.400pt}{5.059pt}}
\put(700.0,468.0){\rule[-0.200pt]{4.818pt}{0.400pt}}
\put(700.0,489.0){\rule[-0.200pt]{4.818pt}{0.400pt}}
\put(760.0,439.0){\rule[-0.200pt]{0.400pt}{8.913pt}}
\put(750.0,439.0){\rule[-0.200pt]{4.818pt}{0.400pt}}
\put(750.0,476.0){\rule[-0.200pt]{4.818pt}{0.400pt}}
\put(811.0,398.0){\rule[-0.200pt]{0.400pt}{10.600pt}}
\put(801.0,398.0){\rule[-0.200pt]{4.818pt}{0.400pt}}
\put(801.0,442.0){\rule[-0.200pt]{4.818pt}{0.400pt}}
\put(861.0,386.0){\rule[-0.200pt]{0.400pt}{2.168pt}}
\put(851.0,386.0){\rule[-0.200pt]{4.818pt}{0.400pt}}
\put(851.0,395.0){\rule[-0.200pt]{4.818pt}{0.400pt}}
\put(911.0,342.0){\rule[-0.200pt]{0.400pt}{2.409pt}}
\put(901.0,342.0){\rule[-0.200pt]{4.818pt}{0.400pt}}
\put(901.0,352.0){\rule[-0.200pt]{4.818pt}{0.400pt}}
\put(962.0,298.0){\rule[-0.200pt]{0.400pt}{4.818pt}}
\put(952.0,298.0){\rule[-0.200pt]{4.818pt}{0.400pt}}
\put(952.0,318.0){\rule[-0.200pt]{4.818pt}{0.400pt}}
\put(1012.0,260.0){\rule[-0.200pt]{0.400pt}{1.445pt}}
\put(1002.0,260.0){\rule[-0.200pt]{4.818pt}{0.400pt}}
\put(1002.0,266.0){\rule[-0.200pt]{4.818pt}{0.400pt}}
\put(1062.0,228.0){\rule[-0.200pt]{0.400pt}{0.723pt}}
\put(1052.0,228.0){\rule[-0.200pt]{4.818pt}{0.400pt}}
\put(1052.0,231.0){\rule[-0.200pt]{4.818pt}{0.400pt}}
\put(1113.0,209.0){\rule[-0.200pt]{0.400pt}{0.482pt}}
\put(1103.0,209.0){\rule[-0.200pt]{4.818pt}{0.400pt}}
\put(1103.0,211.0){\rule[-0.200pt]{4.818pt}{0.400pt}}
\put(1163.0,204.0){\usebox{\plotpoint}}
\put(1153.0,204.0){\rule[-0.200pt]{4.818pt}{0.400pt}}
\put(1153.0,205.0){\rule[-0.200pt]{4.818pt}{0.400pt}}
\put(1213,204){\usebox{\plotpoint}}
\put(1203.0,204.0){\rule[-0.200pt]{4.818pt}{0.400pt}}
\put(1203.0,204.0){\rule[-0.200pt]{4.818pt}{0.400pt}}
\put(1264,204){\usebox{\plotpoint}}
\put(1254.0,204.0){\rule[-0.200pt]{4.818pt}{0.400pt}}
\put(1254.0,204.0){\rule[-0.200pt]{4.818pt}{0.400pt}}
\put(1314,204){\usebox{\plotpoint}}
\put(1304.0,204.0){\rule[-0.200pt]{4.818pt}{0.400pt}}
\put(1304.0,204.0){\rule[-0.200pt]{4.818pt}{0.400pt}}
\put(1364,204){\usebox{\plotpoint}}
\put(1354.0,204.0){\rule[-0.200pt]{4.818pt}{0.400pt}}
\put(1354.0,204.0){\rule[-0.200pt]{4.818pt}{0.400pt}}
\put(1415,204){\usebox{\plotpoint}}
\put(1405.0,204.0){\rule[-0.200pt]{4.818pt}{0.400pt}}
\put(206,540){\raisebox{-.8pt}{\makebox(0,0){$\Diamond$}}}
\put(257,766){\raisebox{-.8pt}{\makebox(0,0){$\Diamond$}}}
\put(307,749){\raisebox{-.8pt}{\makebox(0,0){$\Diamond$}}}
\put(357,702){\raisebox{-.8pt}{\makebox(0,0){$\Diamond$}}}
\put(408,674){\raisebox{-.8pt}{\makebox(0,0){$\Diamond$}}}
\put(458,646){\raisebox{-.8pt}{\makebox(0,0){$\Diamond$}}}
\put(508,602){\raisebox{-.8pt}{\makebox(0,0){$\Diamond$}}}
\put(559,572){\raisebox{-.8pt}{\makebox(0,0){$\Diamond$}}}
\put(609,545){\raisebox{-.8pt}{\makebox(0,0){$\Diamond$}}}
\put(659,514){\raisebox{-.8pt}{\makebox(0,0){$\Diamond$}}}
\put(710,479){\raisebox{-.8pt}{\makebox(0,0){$\Diamond$}}}
\put(760,457){\raisebox{-.8pt}{\makebox(0,0){$\Diamond$}}}
\put(811,420){\raisebox{-.8pt}{\makebox(0,0){$\Diamond$}}}
\put(861,391){\raisebox{-.8pt}{\makebox(0,0){$\Diamond$}}}
\put(911,347){\raisebox{-.8pt}{\makebox(0,0){$\Diamond$}}}
\put(962,308){\raisebox{-.8pt}{\makebox(0,0){$\Diamond$}}}
\put(1012,263){\raisebox{-.8pt}{\makebox(0,0){$\Diamond$}}}
\put(1062,229){\raisebox{-.8pt}{\makebox(0,0){$\Diamond$}}}
\put(1113,210){\raisebox{-.8pt}{\makebox(0,0){$\Diamond$}}}
\put(1163,204){\raisebox{-.8pt}{\makebox(0,0){$\Diamond$}}}
\put(1213,204){\raisebox{-.8pt}{\makebox(0,0){$\Diamond$}}}
\put(1264,204){\raisebox{-.8pt}{\makebox(0,0){$\Diamond$}}}
\put(1314,204){\raisebox{-.8pt}{\makebox(0,0){$\Diamond$}}}
\put(1364,204){\raisebox{-.8pt}{\makebox(0,0){$\Diamond$}}}
\put(1415,204){\raisebox{-.8pt}{\makebox(0,0){$\Diamond$}}}
\put(1405.0,204.0){\rule[-0.200pt]{4.818pt}{0.400pt}}
\multiput(206,785)(0.000,20.756){2}{\usebox{\plotpoint}}
\put(206,821){\usebox{\plotpoint}}
\put(196.00,785.00){\usebox{\plotpoint}}
\put(216,785){\usebox{\plotpoint}}
\put(196.00,821.00){\usebox{\plotpoint}}
\put(216,821){\usebox{\plotpoint}}
\put(257.00,620.00){\usebox{\plotpoint}}
\put(257,628){\usebox{\plotpoint}}
\put(247.00,620.00){\usebox{\plotpoint}}
\put(267,620){\usebox{\plotpoint}}
\put(247.00,628.00){\usebox{\plotpoint}}
\put(267,628){\usebox{\plotpoint}}
\put(307.00,549.00){\usebox{\plotpoint}}
\put(307,559){\usebox{\plotpoint}}
\put(297.00,549.00){\usebox{\plotpoint}}
\put(317,549){\usebox{\plotpoint}}
\put(297.00,559.00){\usebox{\plotpoint}}
\put(317,559){\usebox{\plotpoint}}
\put(357.00,507.00){\usebox{\plotpoint}}
\put(357,516){\usebox{\plotpoint}}
\put(347.00,507.00){\usebox{\plotpoint}}
\put(367,507){\usebox{\plotpoint}}
\put(347.00,516.00){\usebox{\plotpoint}}
\put(367,516){\usebox{\plotpoint}}
\put(408.00,474.00){\usebox{\plotpoint}}
\put(408,484){\usebox{\plotpoint}}
\put(398.00,474.00){\usebox{\plotpoint}}
\put(418,474){\usebox{\plotpoint}}
\put(398.00,484.00){\usebox{\plotpoint}}
\put(418,484){\usebox{\plotpoint}}
\put(458.00,446.00){\usebox{\plotpoint}}
\put(458,451){\usebox{\plotpoint}}
\put(448.00,446.00){\usebox{\plotpoint}}
\put(468,446){\usebox{\plotpoint}}
\put(448.00,451.00){\usebox{\plotpoint}}
\put(468,451){\usebox{\plotpoint}}
\put(508.00,425.00){\usebox{\plotpoint}}
\put(508,431){\usebox{\plotpoint}}
\put(498.00,425.00){\usebox{\plotpoint}}
\put(518,425){\usebox{\plotpoint}}
\put(498.00,431.00){\usebox{\plotpoint}}
\put(518,431){\usebox{\plotpoint}}
\put(559.00,407.00){\usebox{\plotpoint}}
\put(559,410){\usebox{\plotpoint}}
\put(549.00,407.00){\usebox{\plotpoint}}
\put(569,407){\usebox{\plotpoint}}
\put(549.00,410.00){\usebox{\plotpoint}}
\put(569,410){\usebox{\plotpoint}}
\put(609.00,388.00){\usebox{\plotpoint}}
\put(609,393){\usebox{\plotpoint}}
\put(599.00,388.00){\usebox{\plotpoint}}
\put(619,388){\usebox{\plotpoint}}
\put(599.00,393.00){\usebox{\plotpoint}}
\put(619,393){\usebox{\plotpoint}}
\put(659.00,373.00){\usebox{\plotpoint}}
\put(659,375){\usebox{\plotpoint}}
\put(649.00,373.00){\usebox{\plotpoint}}
\put(669,373){\usebox{\plotpoint}}
\put(649.00,375.00){\usebox{\plotpoint}}
\put(669,375){\usebox{\plotpoint}}
\put(710.00,356.00){\usebox{\plotpoint}}
\put(710,358){\usebox{\plotpoint}}
\put(700.00,356.00){\usebox{\plotpoint}}
\put(720,356){\usebox{\plotpoint}}
\put(700.00,358.00){\usebox{\plotpoint}}
\put(720,358){\usebox{\plotpoint}}
\put(760.00,339.00){\usebox{\plotpoint}}
\put(760,342){\usebox{\plotpoint}}
\put(750.00,339.00){\usebox{\plotpoint}}
\put(770,339){\usebox{\plotpoint}}
\put(750.00,342.00){\usebox{\plotpoint}}
\put(770,342){\usebox{\plotpoint}}
\put(811.00,320.00){\usebox{\plotpoint}}
\put(811,324){\usebox{\plotpoint}}
\put(801.00,320.00){\usebox{\plotpoint}}
\put(821,320){\usebox{\plotpoint}}
\put(801.00,324.00){\usebox{\plotpoint}}
\put(821,324){\usebox{\plotpoint}}
\put(861.00,302.00){\usebox{\plotpoint}}
\put(861,303){\usebox{\plotpoint}}
\put(851.00,302.00){\usebox{\plotpoint}}
\put(871,302){\usebox{\plotpoint}}
\put(851.00,303.00){\usebox{\plotpoint}}
\put(871,303){\usebox{\plotpoint}}
\put(911.00,280.00){\usebox{\plotpoint}}
\put(911,282){\usebox{\plotpoint}}
\put(901.00,280.00){\usebox{\plotpoint}}
\put(921,280){\usebox{\plotpoint}}
\put(901.00,282.00){\usebox{\plotpoint}}
\put(921,282){\usebox{\plotpoint}}
\put(962.00,256.00){\usebox{\plotpoint}}
\put(962,258){\usebox{\plotpoint}}
\put(952.00,256.00){\usebox{\plotpoint}}
\put(972,256){\usebox{\plotpoint}}
\put(952.00,258.00){\usebox{\plotpoint}}
\put(972,258){\usebox{\plotpoint}}
\put(1012.00,234.00){\usebox{\plotpoint}}
\put(1012,235){\usebox{\plotpoint}}
\put(1002.00,234.00){\usebox{\plotpoint}}
\put(1022,234){\usebox{\plotpoint}}
\put(1002.00,235.00){\usebox{\plotpoint}}
\put(1022,235){\usebox{\plotpoint}}
\put(1062,217){\usebox{\plotpoint}}
\put(1062,217){\usebox{\plotpoint}}
\put(1052.00,217.00){\usebox{\plotpoint}}
\put(1072,217){\usebox{\plotpoint}}
\put(1052.00,217.00){\usebox{\plotpoint}}
\put(1072,217){\usebox{\plotpoint}}
\put(1113.00,206.00){\usebox{\plotpoint}}
\put(1113,207){\usebox{\plotpoint}}
\put(1103.00,206.00){\usebox{\plotpoint}}
\put(1123,206){\usebox{\plotpoint}}
\put(1103.00,207.00){\usebox{\plotpoint}}
\put(1123,207){\usebox{\plotpoint}}
\put(1163,204){\usebox{\plotpoint}}
\put(1163,204){\usebox{\plotpoint}}
\put(1153.00,204.00){\usebox{\plotpoint}}
\put(1173,204){\usebox{\plotpoint}}
\put(1153.00,204.00){\usebox{\plotpoint}}
\put(1173,204){\usebox{\plotpoint}}
\put(1213,204){\usebox{\plotpoint}}
\put(1213,204){\usebox{\plotpoint}}
\put(1203.00,204.00){\usebox{\plotpoint}}
\put(1223,204){\usebox{\plotpoint}}
\put(1203.00,204.00){\usebox{\plotpoint}}
\put(1223,204){\usebox{\plotpoint}}
\put(1264,204){\usebox{\plotpoint}}
\put(1264,204){\usebox{\plotpoint}}
\put(1254.00,204.00){\usebox{\plotpoint}}
\put(1274,204){\usebox{\plotpoint}}
\put(1254.00,204.00){\usebox{\plotpoint}}
\put(1274,204){\usebox{\plotpoint}}
\put(1314,204){\usebox{\plotpoint}}
\put(1314,204){\usebox{\plotpoint}}
\put(1304.00,204.00){\usebox{\plotpoint}}
\put(1324,204){\usebox{\plotpoint}}
\put(1304.00,204.00){\usebox{\plotpoint}}
\put(1324,204){\usebox{\plotpoint}}
\put(1364,204){\usebox{\plotpoint}}
\put(1364,204){\usebox{\plotpoint}}
\put(1354.00,204.00){\usebox{\plotpoint}}
\put(1374,204){\usebox{\plotpoint}}
\put(1354.00,204.00){\usebox{\plotpoint}}
\put(1374,204){\usebox{\plotpoint}}
\put(1415,204){\usebox{\plotpoint}}
\put(1415,204){\usebox{\plotpoint}}
\put(1405.00,204.00){\usebox{\plotpoint}}
\put(1425,204){\usebox{\plotpoint}}
\put(1405.00,204.00){\usebox{\plotpoint}}
\put(1425,204){\usebox{\plotpoint}}
\put(206,803){\makebox(0,0){$+$}}
\put(257,624){\makebox(0,0){$+$}}
\put(307,554){\makebox(0,0){$+$}}
\put(357,511){\makebox(0,0){$+$}}
\put(408,479){\makebox(0,0){$+$}}
\put(458,448){\makebox(0,0){$+$}}
\put(508,428){\makebox(0,0){$+$}}
\put(559,409){\makebox(0,0){$+$}}
\put(609,391){\makebox(0,0){$+$}}
\put(659,374){\makebox(0,0){$+$}}
\put(710,357){\makebox(0,0){$+$}}
\put(760,341){\makebox(0,0){$+$}}
\put(811,322){\makebox(0,0){$+$}}
\put(861,302){\makebox(0,0){$+$}}
\put(911,281){\makebox(0,0){$+$}}
\put(962,257){\makebox(0,0){$+$}}
\put(1012,235){\makebox(0,0){$+$}}
\put(1062,217){\makebox(0,0){$+$}}
\put(1113,207){\makebox(0,0){$+$}}
\put(1163,204){\makebox(0,0){$+$}}
\put(1213,204){\makebox(0,0){$+$}}
\put(1264,204){\makebox(0,0){$+$}}
\put(1314,204){\makebox(0,0){$+$}}
\put(1364,204){\makebox(0,0){$+$}}
\put(1415,204){\makebox(0,0){$+$}}
\sbox{\plotpoint}{\rule[-0.400pt]{0.800pt}{0.800pt}}%
\put(181,204){\usebox{\plotpoint}}
\put(181.0,204.0){\rule[-0.400pt]{303.293pt}{0.800pt}}
\end{picture}
\caption{The $B_{jet}$ distribution at NLO (diamonds) and LO (crosses).}
\end{center}
\end{figure}
\begin{table}
\begin{center}
\begin{tabular}{|c|cc|} \hline
$B_{jet}$ & $B_{B_{jet}}$ & $C_{B_{jet}}$ \\ \hline
0.02 & $ ( 7.31 \pm 0.22 ) \cdot 10^{1} $ & $(-1.71 \pm 0.66) \cdot 10^{3}$ \\
0.06 & $ ( 5.13 \pm 0.05 ) \cdot 10^{1} $ & $( 9.23 \pm 2.79) \cdot 10^{2}$ \\
0.10 & $ ( 4.27 \pm 0.06 ) \cdot 10^{1} $ & $( 1.27 \pm 0.20) \cdot 10^{3}$ \\
0.14 & $ ( 3.75 \pm 0.06 ) \cdot 10^{1} $ & $( 1.24 \pm 0.16) \cdot 10^{3}$ \\
0.18 & $ ( 3.36 \pm 0.06 ) \cdot 10^{1} $ & $( 1.26 \pm 0.14) \cdot 10^{3}$\\
0.22 & $ ( 2.99 \pm 0.03 ) \cdot 10^{1} $ & $( 1.28 \pm 0.14) \cdot 10^{3}$\\
0.26 & $ ( 2.74 \pm 0.03 ) \cdot 10^{1} $ & $( 1.13 \pm 0.12) \cdot 10^{3}$\\
0.30 & $ ( 2.50 \pm 0.02 ) \cdot 10^{1} $ & $( 1.06 \pm 0.07) \cdot 10^{3}$\\
0.34 & $ ( 2.28 \pm 0.03 ) \cdot 10^{1} $ & $( 1.00 \pm 0.07) \cdot 10^{3}$ \\
0.38 & $ ( 2.08 \pm 0.02 ) \cdot 10^{1} $ & $( 9.12 \pm 0.51) \cdot 10^{2}$ \\
0.42 & $ ( 1.87 \pm 0.01 ) \cdot 10^{1} $ & $( 7.92 \pm 0.69) \cdot 10^{2}$ \\
0.46 & $ ( 1.67 \pm 0.02 ) \cdot 10^{1} $ & $( 7.59 \pm 1.19) \cdot 10^{2}$ \\
0.50 & $ ( 1.44 \pm 0.02 ) \cdot 10^{1} $ & $( 6.36 \pm 1.43) \cdot 10^{2}$ \\
0.54 & $ ( 1.20 \pm 0.01 ) \cdot 10^{1} $ & $( 5.75 \pm 0.30) \cdot 10^{2}$ \\
0.58 & $ ( 9.42 \pm 0.18 ) \cdot 10^{0} $ & $( 4.30 \pm 0.33) \cdot 10^{2}$ \\
0.62 & $ ( 6.46 \pm 0.12 ) \cdot 10^{0} $ & $( 3.33 \pm 0.65) \cdot 10^{2}$ \\
0.66 & $ ( 3.76 \pm 0.07 ) \cdot 10^{0} $ & $( 1.84 \pm 0.18) \cdot 10^{2}$ \\
0.70 & $ ( 1.60 \pm 0.04 ) \cdot 10^{0} $ & $( 7.97 \pm 1.07) \cdot 10^{1}$ \\
0.74 & $ ( 3.39 \pm 0.28 ) \cdot 10^{-1} $ & $( 1.95 \pm 0.55) \cdot 10^{1}$ \\
0.78 & $ ( 1.11 \pm 0.31 ) \cdot 10^{-2} $ & $( 2.79 \pm 1.24) \cdot 10^{0}$ \\
0.82 & $ 0.00 $ & $( 8.26 \pm 9.73) \cdot 10^{-2}$ \\
0.86 & $ 0.00 $ & $ 0.00 $ \\
0.90 & $ 0.00 $ & $ 0.00 $ \\
0.94 & $ 0.00 $ & $ 0.00 $ \\
0.98 & $ 0.00 $ & $ 0.00 $ \\ \hline
\end{tabular}
\end{center}
\caption{The Born level and next-to-leading order functions $B_{B_{jet}}$ and $C_{B_{jet}}$ for the jet broadening variable.}
\end{table}
The values for the functions $B_{B_{jet}}$ and $C_{B_{jet}}$ are given in table 3.
Figure 2 shows the distribution of the jet broadening variable.

\subsection{The Softest-jet Explanarity}

For three jet events we define the softest-jet explanarity as follows:
\begin{eqnarray}
E_{3 jet} & = & \frac{\sum\limits_{a} \left| \vec{p}_a \cdot \left( \vec{P}_1 \times \vec{P}_2 \right) \right|}{\sum\limits_{a} \left| \vec{p}_a \right| \left| \vec{P}_1 \times \vec{P}_2 \right|}
\end{eqnarray}
where $\vec{p}_a$ is the three-momentum of particle $a$ inside the softest jet and the sum runs over all particles in the softest jet. $\vec{P}_1$ and $\vec{P}_2$ denote the three-momenta of the remaining two jets in the event.
The explanarity measures the degree to which the shape of the softest jet lies out of the event plane.
The definition is motivated by the fact, that the explanarity will be sensitive to the color factors of QCD.\\
\\
The softest-jet explanarity is calculated for three-jet events defined by the DURHAM algorithm and $y_{cut} = 0.1$.
For the average we obtain
\begin{eqnarray}
\langle E_{3 jet} \rangle & = & \left( \frac{\alpha_s}{2 \pi} \right)^2 \left( 1.29 \pm 0.01 \right) \cdot 10^{1}
                             +\left( \frac{\alpha_s}{2 \pi} \right)^3 \left( 3.33 \pm 0.22 \right) \cdot 10^{2}.
\end{eqnarray}
\begin{figure}
\begin{center}
%\input fig_3_jet.tex
% GNUPLOT: LaTeX picture
\setlength{\unitlength}{0.240900pt}
\ifx\plotpoint\undefined\newsavebox{\plotpoint}\fi
\sbox{\plotpoint}{\rule[-0.200pt]{0.400pt}{0.400pt}}%
\begin{picture}(1500,900)(0,0)
\font\gnuplot=cmr10 at 10pt
\gnuplot
\sbox{\plotpoint}{\rule[-0.200pt]{0.400pt}{0.400pt}}%
\put(181.0,122.0){\rule[-0.200pt]{4.818pt}{0.400pt}}
\put(161,122){\makebox(0,0)[r]{-10}}
\put(1420.0,122.0){\rule[-0.200pt]{4.818pt}{0.400pt}}
\put(181.0,204.0){\rule[-0.200pt]{4.818pt}{0.400pt}}
\put(161,204){\makebox(0,0)[r]{0}}
\put(1420.0,204.0){\rule[-0.200pt]{4.818pt}{0.400pt}}
\put(181.0,286.0){\rule[-0.200pt]{4.818pt}{0.400pt}}
\put(161,286){\makebox(0,0)[r]{10}}
\put(1420.0,286.0){\rule[-0.200pt]{4.818pt}{0.400pt}}
\put(181.0,368.0){\rule[-0.200pt]{4.818pt}{0.400pt}}
\put(161,368){\makebox(0,0)[r]{20}}
\put(1420.0,368.0){\rule[-0.200pt]{4.818pt}{0.400pt}}
\put(181.0,450.0){\rule[-0.200pt]{4.818pt}{0.400pt}}
\put(161,450){\makebox(0,0)[r]{30}}
\put(1420.0,450.0){\rule[-0.200pt]{4.818pt}{0.400pt}}
\put(181.0,531.0){\rule[-0.200pt]{4.818pt}{0.400pt}}
\put(161,531){\makebox(0,0)[r]{40}}
\put(1420.0,531.0){\rule[-0.200pt]{4.818pt}{0.400pt}}
\put(181.0,613.0){\rule[-0.200pt]{4.818pt}{0.400pt}}
\put(161,613){\makebox(0,0)[r]{50}}
\put(1420.0,613.0){\rule[-0.200pt]{4.818pt}{0.400pt}}
\put(181.0,695.0){\rule[-0.200pt]{4.818pt}{0.400pt}}
\put(161,695){\makebox(0,0)[r]{60}}
\put(1420.0,695.0){\rule[-0.200pt]{4.818pt}{0.400pt}}
\put(181.0,777.0){\rule[-0.200pt]{4.818pt}{0.400pt}}
\put(161,777){\makebox(0,0)[r]{70}}
\put(1420.0,777.0){\rule[-0.200pt]{4.818pt}{0.400pt}}
\put(181.0,859.0){\rule[-0.200pt]{4.818pt}{0.400pt}}
\put(161,859){\makebox(0,0)[r]{80}}
\put(1420.0,859.0){\rule[-0.200pt]{4.818pt}{0.400pt}}
\put(181.0,122.0){\rule[-0.200pt]{0.400pt}{4.818pt}}
\put(181,81){\makebox(0,0){0}}
\put(181.0,839.0){\rule[-0.200pt]{0.400pt}{4.818pt}}
\put(433.0,122.0){\rule[-0.200pt]{0.400pt}{4.818pt}}
\put(433,81){\makebox(0,0){0.2}}
\put(433.0,839.0){\rule[-0.200pt]{0.400pt}{4.818pt}}
\put(685.0,122.0){\rule[-0.200pt]{0.400pt}{4.818pt}}
\put(685,81){\makebox(0,0){0.4}}
\put(685.0,839.0){\rule[-0.200pt]{0.400pt}{4.818pt}}
\put(936.0,122.0){\rule[-0.200pt]{0.400pt}{4.818pt}}
\put(936,81){\makebox(0,0){0.6}}
\put(936.0,839.0){\rule[-0.200pt]{0.400pt}{4.818pt}}
\put(1188.0,122.0){\rule[-0.200pt]{0.400pt}{4.818pt}}
\put(1188,81){\makebox(0,0){0.8}}
\put(1188.0,839.0){\rule[-0.200pt]{0.400pt}{4.818pt}}
\put(1440.0,122.0){\rule[-0.200pt]{0.400pt}{4.818pt}}
\put(1440,81){\makebox(0,0){1}}
\put(1440.0,839.0){\rule[-0.200pt]{0.400pt}{4.818pt}}
\put(181.0,122.0){\rule[-0.200pt]{303.293pt}{0.400pt}}
\put(1440.0,122.0){\rule[-0.200pt]{0.400pt}{177.543pt}}
\put(181.0,859.0){\rule[-0.200pt]{303.293pt}{0.400pt}}
\put(41,490){\makebox(0,0){\shortstack{$\left( \frac{\alpha_s}{2 \pi} \right)^{-2} \frac{E_{3 jet}}{\sigma_{tot}} \;\;\;\;\;\;\;$ \\ $ \cdot \frac{d \sigma}{d E_{3 jet}}$}}}
\put(810,40){\makebox(0,0){$E_{3 jet}$}}
\put(181.0,122.0){\rule[-0.200pt]{0.400pt}{177.543pt}}
\put(206.0,497.0){\rule[-0.200pt]{0.400pt}{43.603pt}}
\put(196.0,497.0){\rule[-0.200pt]{4.818pt}{0.400pt}}
\put(196.0,678.0){\rule[-0.200pt]{4.818pt}{0.400pt}}
\put(257.0,758.0){\rule[-0.200pt]{0.400pt}{9.877pt}}
\put(247.0,758.0){\rule[-0.200pt]{4.818pt}{0.400pt}}
\put(247.0,799.0){\rule[-0.200pt]{4.818pt}{0.400pt}}
\put(307.0,683.0){\rule[-0.200pt]{0.400pt}{11.804pt}}
\put(297.0,683.0){\rule[-0.200pt]{4.818pt}{0.400pt}}
\put(297.0,732.0){\rule[-0.200pt]{4.818pt}{0.400pt}}
\put(357.0,613.0){\rule[-0.200pt]{0.400pt}{8.672pt}}
\put(347.0,613.0){\rule[-0.200pt]{4.818pt}{0.400pt}}
\put(347.0,649.0){\rule[-0.200pt]{4.818pt}{0.400pt}}
\put(408.0,549.0){\rule[-0.200pt]{0.400pt}{8.191pt}}
\put(398.0,549.0){\rule[-0.200pt]{4.818pt}{0.400pt}}
\put(398.0,583.0){\rule[-0.200pt]{4.818pt}{0.400pt}}
\put(458.0,514.0){\rule[-0.200pt]{0.400pt}{6.263pt}}
\put(448.0,514.0){\rule[-0.200pt]{4.818pt}{0.400pt}}
\put(448.0,540.0){\rule[-0.200pt]{4.818pt}{0.400pt}}
\put(508.0,458.0){\rule[-0.200pt]{0.400pt}{10.118pt}}
\put(498.0,458.0){\rule[-0.200pt]{4.818pt}{0.400pt}}
\put(498.0,500.0){\rule[-0.200pt]{4.818pt}{0.400pt}}
\put(559.0,423.0){\rule[-0.200pt]{0.400pt}{5.300pt}}
\put(549.0,423.0){\rule[-0.200pt]{4.818pt}{0.400pt}}
\put(549.0,445.0){\rule[-0.200pt]{4.818pt}{0.400pt}}
\put(609.0,384.0){\rule[-0.200pt]{0.400pt}{4.577pt}}
\put(599.0,384.0){\rule[-0.200pt]{4.818pt}{0.400pt}}
\put(599.0,403.0){\rule[-0.200pt]{4.818pt}{0.400pt}}
\put(659.0,364.0){\rule[-0.200pt]{0.400pt}{3.373pt}}
\put(649.0,364.0){\rule[-0.200pt]{4.818pt}{0.400pt}}
\put(649.0,378.0){\rule[-0.200pt]{4.818pt}{0.400pt}}
\put(710.0,329.0){\rule[-0.200pt]{0.400pt}{3.854pt}}
\put(700.0,329.0){\rule[-0.200pt]{4.818pt}{0.400pt}}
\put(700.0,345.0){\rule[-0.200pt]{4.818pt}{0.400pt}}
\put(760.0,309.0){\rule[-0.200pt]{0.400pt}{2.650pt}}
\put(750.0,309.0){\rule[-0.200pt]{4.818pt}{0.400pt}}
\put(750.0,320.0){\rule[-0.200pt]{4.818pt}{0.400pt}}
\put(811.0,279.0){\rule[-0.200pt]{0.400pt}{2.650pt}}
\put(801.0,279.0){\rule[-0.200pt]{4.818pt}{0.400pt}}
\put(801.0,290.0){\rule[-0.200pt]{4.818pt}{0.400pt}}
\put(861.0,263.0){\rule[-0.200pt]{0.400pt}{2.409pt}}
\put(851.0,263.0){\rule[-0.200pt]{4.818pt}{0.400pt}}
\put(851.0,273.0){\rule[-0.200pt]{4.818pt}{0.400pt}}
\put(911.0,242.0){\rule[-0.200pt]{0.400pt}{3.854pt}}
\put(901.0,242.0){\rule[-0.200pt]{4.818pt}{0.400pt}}
\put(901.0,258.0){\rule[-0.200pt]{4.818pt}{0.400pt}}
\put(962.0,227.0){\rule[-0.200pt]{0.400pt}{2.168pt}}
\put(952.0,227.0){\rule[-0.200pt]{4.818pt}{0.400pt}}
\put(952.0,236.0){\rule[-0.200pt]{4.818pt}{0.400pt}}
\put(1012.0,216.0){\rule[-0.200pt]{0.400pt}{1.204pt}}
\put(1002.0,216.0){\rule[-0.200pt]{4.818pt}{0.400pt}}
\put(1002.0,221.0){\rule[-0.200pt]{4.818pt}{0.400pt}}
\put(1062.0,208.0){\rule[-0.200pt]{0.400pt}{0.723pt}}
\put(1052.0,208.0){\rule[-0.200pt]{4.818pt}{0.400pt}}
\put(1052.0,211.0){\rule[-0.200pt]{4.818pt}{0.400pt}}
\put(1113.0,204.0){\usebox{\plotpoint}}
\put(1103.0,204.0){\rule[-0.200pt]{4.818pt}{0.400pt}}
\put(1103.0,205.0){\rule[-0.200pt]{4.818pt}{0.400pt}}
\put(1163,204){\usebox{\plotpoint}}
\put(1153.0,204.0){\rule[-0.200pt]{4.818pt}{0.400pt}}
\put(1153.0,204.0){\rule[-0.200pt]{4.818pt}{0.400pt}}
\put(1213,204){\usebox{\plotpoint}}
\put(1203.0,204.0){\rule[-0.200pt]{4.818pt}{0.400pt}}
\put(1203.0,204.0){\rule[-0.200pt]{4.818pt}{0.400pt}}
\put(1264,204){\usebox{\plotpoint}}
\put(1254.0,204.0){\rule[-0.200pt]{4.818pt}{0.400pt}}
\put(1254.0,204.0){\rule[-0.200pt]{4.818pt}{0.400pt}}
\put(1314,204){\usebox{\plotpoint}}
\put(1304.0,204.0){\rule[-0.200pt]{4.818pt}{0.400pt}}
\put(1304.0,204.0){\rule[-0.200pt]{4.818pt}{0.400pt}}
\put(1364,204){\usebox{\plotpoint}}
\put(1354.0,204.0){\rule[-0.200pt]{4.818pt}{0.400pt}}
\put(1354.0,204.0){\rule[-0.200pt]{4.818pt}{0.400pt}}
\put(1415,204){\usebox{\plotpoint}}
\put(1405.0,204.0){\rule[-0.200pt]{4.818pt}{0.400pt}}
\put(206,588){\raisebox{-.8pt}{\makebox(0,0){$\Diamond$}}}
\put(257,779){\raisebox{-.8pt}{\makebox(0,0){$\Diamond$}}}
\put(307,707){\raisebox{-.8pt}{\makebox(0,0){$\Diamond$}}}
\put(357,631){\raisebox{-.8pt}{\makebox(0,0){$\Diamond$}}}
\put(408,566){\raisebox{-.8pt}{\makebox(0,0){$\Diamond$}}}
\put(458,527){\raisebox{-.8pt}{\makebox(0,0){$\Diamond$}}}
\put(508,479){\raisebox{-.8pt}{\makebox(0,0){$\Diamond$}}}
\put(559,434){\raisebox{-.8pt}{\makebox(0,0){$\Diamond$}}}
\put(609,394){\raisebox{-.8pt}{\makebox(0,0){$\Diamond$}}}
\put(659,371){\raisebox{-.8pt}{\makebox(0,0){$\Diamond$}}}
\put(710,337){\raisebox{-.8pt}{\makebox(0,0){$\Diamond$}}}
\put(760,315){\raisebox{-.8pt}{\makebox(0,0){$\Diamond$}}}
\put(811,285){\raisebox{-.8pt}{\makebox(0,0){$\Diamond$}}}
\put(861,268){\raisebox{-.8pt}{\makebox(0,0){$\Diamond$}}}
\put(911,250){\raisebox{-.8pt}{\makebox(0,0){$\Diamond$}}}
\put(962,232){\raisebox{-.8pt}{\makebox(0,0){$\Diamond$}}}
\put(1012,219){\raisebox{-.8pt}{\makebox(0,0){$\Diamond$}}}
\put(1062,210){\raisebox{-.8pt}{\makebox(0,0){$\Diamond$}}}
\put(1113,204){\raisebox{-.8pt}{\makebox(0,0){$\Diamond$}}}
\put(1163,204){\raisebox{-.8pt}{\makebox(0,0){$\Diamond$}}}
\put(1213,204){\raisebox{-.8pt}{\makebox(0,0){$\Diamond$}}}
\put(1264,204){\raisebox{-.8pt}{\makebox(0,0){$\Diamond$}}}
\put(1314,204){\raisebox{-.8pt}{\makebox(0,0){$\Diamond$}}}
\put(1364,204){\raisebox{-.8pt}{\makebox(0,0){$\Diamond$}}}
\put(1415,204){\raisebox{-.8pt}{\makebox(0,0){$\Diamond$}}}
\put(1405.0,204.0){\rule[-0.200pt]{4.818pt}{0.400pt}}
\multiput(206,769)(0.000,20.756){2}{\usebox{\plotpoint}}
\put(206,795){\usebox{\plotpoint}}
\put(196.00,769.00){\usebox{\plotpoint}}
\put(216,769){\usebox{\plotpoint}}
\put(196.00,795.00){\usebox{\plotpoint}}
\put(216,795){\usebox{\plotpoint}}
\put(257.00,574.00){\usebox{\plotpoint}}
\put(257,585){\usebox{\plotpoint}}
\put(247.00,574.00){\usebox{\plotpoint}}
\put(267,574){\usebox{\plotpoint}}
\put(247.00,585.00){\usebox{\plotpoint}}
\put(267,585){\usebox{\plotpoint}}
\put(307.00,496.00){\usebox{\plotpoint}}
\put(307,503){\usebox{\plotpoint}}
\put(297.00,496.00){\usebox{\plotpoint}}
\put(317,496){\usebox{\plotpoint}}
\put(297.00,503.00){\usebox{\plotpoint}}
\put(317,503){\usebox{\plotpoint}}
\put(357.00,447.00){\usebox{\plotpoint}}
\put(357,451){\usebox{\plotpoint}}
\put(347.00,447.00){\usebox{\plotpoint}}
\put(367,447){\usebox{\plotpoint}}
\put(347.00,451.00){\usebox{\plotpoint}}
\put(367,451){\usebox{\plotpoint}}
\put(408.00,408.00){\usebox{\plotpoint}}
\put(408,411){\usebox{\plotpoint}}
\put(398.00,408.00){\usebox{\plotpoint}}
\put(418,408){\usebox{\plotpoint}}
\put(398.00,411.00){\usebox{\plotpoint}}
\put(418,411){\usebox{\plotpoint}}
\put(458.00,377.00){\usebox{\plotpoint}}
\put(458,380){\usebox{\plotpoint}}
\put(448.00,377.00){\usebox{\plotpoint}}
\put(468,377){\usebox{\plotpoint}}
\put(448.00,380.00){\usebox{\plotpoint}}
\put(468,380){\usebox{\plotpoint}}
\put(508.00,353.00){\usebox{\plotpoint}}
\put(508,355){\usebox{\plotpoint}}
\put(498.00,353.00){\usebox{\plotpoint}}
\put(518,353){\usebox{\plotpoint}}
\put(498.00,355.00){\usebox{\plotpoint}}
\put(518,355){\usebox{\plotpoint}}
\put(559.00,329.00){\usebox{\plotpoint}}
\put(559,333){\usebox{\plotpoint}}
\put(549.00,329.00){\usebox{\plotpoint}}
\put(569,329){\usebox{\plotpoint}}
\put(549.00,333.00){\usebox{\plotpoint}}
\put(569,333){\usebox{\plotpoint}}
\put(609.00,311.00){\usebox{\plotpoint}}
\put(609,314){\usebox{\plotpoint}}
\put(599.00,311.00){\usebox{\plotpoint}}
\put(619,311){\usebox{\plotpoint}}
\put(599.00,314.00){\usebox{\plotpoint}}
\put(619,314){\usebox{\plotpoint}}
\put(659.00,294.00){\usebox{\plotpoint}}
\put(659,297){\usebox{\plotpoint}}
\put(649.00,294.00){\usebox{\plotpoint}}
\put(669,294){\usebox{\plotpoint}}
\put(649.00,297.00){\usebox{\plotpoint}}
\put(669,297){\usebox{\plotpoint}}
\put(710.00,279.00){\usebox{\plotpoint}}
\put(710,281){\usebox{\plotpoint}}
\put(700.00,279.00){\usebox{\plotpoint}}
\put(720,279){\usebox{\plotpoint}}
\put(700.00,281.00){\usebox{\plotpoint}}
\put(720,281){\usebox{\plotpoint}}
\put(760.00,265.00){\usebox{\plotpoint}}
\put(760,267){\usebox{\plotpoint}}
\put(750.00,265.00){\usebox{\plotpoint}}
\put(770,265){\usebox{\plotpoint}}
\put(750.00,267.00){\usebox{\plotpoint}}
\put(770,267){\usebox{\plotpoint}}
\put(811.00,252.00){\usebox{\plotpoint}}
\put(811,255){\usebox{\plotpoint}}
\put(801.00,252.00){\usebox{\plotpoint}}
\put(821,252){\usebox{\plotpoint}}
\put(801.00,255.00){\usebox{\plotpoint}}
\put(821,255){\usebox{\plotpoint}}
\put(861.00,241.00){\usebox{\plotpoint}}
\put(861,242){\usebox{\plotpoint}}
\put(851.00,241.00){\usebox{\plotpoint}}
\put(871,241){\usebox{\plotpoint}}
\put(851.00,242.00){\usebox{\plotpoint}}
\put(871,242){\usebox{\plotpoint}}
\put(911.00,232.00){\usebox{\plotpoint}}
\put(911,233){\usebox{\plotpoint}}
\put(901.00,232.00){\usebox{\plotpoint}}
\put(921,232){\usebox{\plotpoint}}
\put(901.00,233.00){\usebox{\plotpoint}}
\put(921,233){\usebox{\plotpoint}}
\put(962.00,222.00){\usebox{\plotpoint}}
\put(962,223){\usebox{\plotpoint}}
\put(952.00,222.00){\usebox{\plotpoint}}
\put(972,222){\usebox{\plotpoint}}
\put(952.00,223.00){\usebox{\plotpoint}}
\put(972,223){\usebox{\plotpoint}}
\put(1012.00,214.00){\usebox{\plotpoint}}
\put(1012,215){\usebox{\plotpoint}}
\put(1002.00,214.00){\usebox{\plotpoint}}
\put(1022,214){\usebox{\plotpoint}}
\put(1002.00,215.00){\usebox{\plotpoint}}
\put(1022,215){\usebox{\plotpoint}}
\put(1062,208){\usebox{\plotpoint}}
\put(1062,208){\usebox{\plotpoint}}
\put(1052.00,208.00){\usebox{\plotpoint}}
\put(1072,208){\usebox{\plotpoint}}
\put(1052.00,208.00){\usebox{\plotpoint}}
\put(1072,208){\usebox{\plotpoint}}
\put(1113.00,204.00){\usebox{\plotpoint}}
\put(1113,205){\usebox{\plotpoint}}
\put(1103.00,204.00){\usebox{\plotpoint}}
\put(1123,204){\usebox{\plotpoint}}
\put(1103.00,205.00){\usebox{\plotpoint}}
\put(1123,205){\usebox{\plotpoint}}
\put(1163,204){\usebox{\plotpoint}}
\put(1163,204){\usebox{\plotpoint}}
\put(1153.00,204.00){\usebox{\plotpoint}}
\put(1173,204){\usebox{\plotpoint}}
\put(1153.00,204.00){\usebox{\plotpoint}}
\put(1173,204){\usebox{\plotpoint}}
\put(1213,204){\usebox{\plotpoint}}
\put(1213,204){\usebox{\plotpoint}}
\put(1203.00,204.00){\usebox{\plotpoint}}
\put(1223,204){\usebox{\plotpoint}}
\put(1203.00,204.00){\usebox{\plotpoint}}
\put(1223,204){\usebox{\plotpoint}}
\put(1264,204){\usebox{\plotpoint}}
\put(1264,204){\usebox{\plotpoint}}
\put(1254.00,204.00){\usebox{\plotpoint}}
\put(1274,204){\usebox{\plotpoint}}
\put(1254.00,204.00){\usebox{\plotpoint}}
\put(1274,204){\usebox{\plotpoint}}
\put(1314,204){\usebox{\plotpoint}}
\put(1314,204){\usebox{\plotpoint}}
\put(1304.00,204.00){\usebox{\plotpoint}}
\put(1324,204){\usebox{\plotpoint}}
\put(1304.00,204.00){\usebox{\plotpoint}}
\put(1324,204){\usebox{\plotpoint}}
\put(1364,204){\usebox{\plotpoint}}
\put(1364,204){\usebox{\plotpoint}}
\put(1354.00,204.00){\usebox{\plotpoint}}
\put(1374,204){\usebox{\plotpoint}}
\put(1354.00,204.00){\usebox{\plotpoint}}
\put(1374,204){\usebox{\plotpoint}}
\put(1415,204){\usebox{\plotpoint}}
\put(1415,204){\usebox{\plotpoint}}
\put(1405.00,204.00){\usebox{\plotpoint}}
\put(1425,204){\usebox{\plotpoint}}
\put(1405.00,204.00){\usebox{\plotpoint}}
\put(1425,204){\usebox{\plotpoint}}
\put(206,782){\makebox(0,0){$+$}}
\put(257,579){\makebox(0,0){$+$}}
\put(307,500){\makebox(0,0){$+$}}
\put(357,449){\makebox(0,0){$+$}}
\put(408,410){\makebox(0,0){$+$}}
\put(458,379){\makebox(0,0){$+$}}
\put(508,354){\makebox(0,0){$+$}}
\put(559,331){\makebox(0,0){$+$}}
\put(609,312){\makebox(0,0){$+$}}
\put(659,296){\makebox(0,0){$+$}}
\put(710,280){\makebox(0,0){$+$}}
\put(760,266){\makebox(0,0){$+$}}
\put(811,254){\makebox(0,0){$+$}}
\put(861,242){\makebox(0,0){$+$}}
\put(911,232){\makebox(0,0){$+$}}
\put(962,222){\makebox(0,0){$+$}}
\put(1012,214){\makebox(0,0){$+$}}
\put(1062,208){\makebox(0,0){$+$}}
\put(1113,204){\makebox(0,0){$+$}}
\put(1163,204){\makebox(0,0){$+$}}
\put(1213,204){\makebox(0,0){$+$}}
\put(1264,204){\makebox(0,0){$+$}}
\put(1314,204){\makebox(0,0){$+$}}
\put(1364,204){\makebox(0,0){$+$}}
\put(1415,204){\makebox(0,0){$+$}}
\sbox{\plotpoint}{\rule[-0.400pt]{0.800pt}{0.800pt}}%
\put(181,204){\usebox{\plotpoint}}
\put(181.0,204.0){\rule[-0.400pt]{303.293pt}{0.800pt}}
\end{picture}
\caption{The $E_{3 jet}$ distribution at NLO (diamonds) and LO (crosses).}
\end{center}
\end{figure}
\begin{table}
\begin{center}
\begin{tabular}{|c|cc|} \hline
$E_{3 jet}$ & $B_{E_{3 jet}}$ & $C_{E_{3 jet}}$ \\ \hline
0.02 & $ ( 7.06 \pm 0.16 ) \cdot 10^{1} $ & $( -1.26 \pm 0.58 ) \cdot 10^{3}$ \\
0.06 & $ ( 4.59 \pm 0.07 ) \cdot 10^{1} $ & $( 1.30 \pm 0.13 ) \cdot 10^{3}$ \\
0.10 & $ ( 3.61 \pm 0.04 ) \cdot 10^{1} $ & $( 1.35 \pm 0.16 ) \cdot 10^{3}$ \\
0.14 & $ ( 2.99 \pm 0.02 ) \cdot 10^{1} $ & $( 1.19 \pm 0.12 ) \cdot 10^{3}$ \\
0.18 & $ ( 2.51 \pm 0.02 ) \cdot 10^{1} $ & $( 1.02 \pm 0.11 ) \cdot 10^{3}$\\
0.22 & $ ( 2.13 \pm 0.02 ) \cdot 10^{1} $ & $( 9.65 \pm 0.86 ) \cdot 10^{2}$\\
0.26 & $ ( 1.83 \pm 0.01 ) \cdot 10^{1} $ & $( 8.13 \pm 1.37 ) \cdot 10^{2}$\\
0.30 & $ ( 1.55 \pm 0.02 ) \cdot 10^{1} $ & $( 6.68 \pm 0.72 ) \cdot 10^{2}$\\
0.34 & $ ( 1.32 \pm 0.02 ) \cdot 10^{1} $ & $( 5.28 \pm 0.60 ) \cdot 10^{2}$ \\
0.38 & $ ( 1.12 \pm 0.01 ) \cdot 10^{1} $ & $( 4.88 \pm 0.45 ) \cdot 10^{2}$ \\
0.42 & $ ( 9.32 \pm 0.13 ) \cdot 10^{0} $ & $( 3.70 \pm 0.53 ) \cdot 10^{2}$ \\
0.46 & $ ( 7.62 \pm 0.10 ) \cdot 10^{0} $ & $( 3.14 \pm 0.33 ) \cdot 10^{2}$ \\
0.50 & $ ( 6.07 \pm 0.15 ) \cdot 10^{0} $ & $( 2.03 \pm 0.34 ) \cdot 10^{2}$ \\
0.54 & $ ( 4.61 \pm 0.10 ) \cdot 10^{0} $ & $( 1.70 \pm 0.32 ) \cdot 10^{2}$ \\
0.58 & $ ( 3.45 \pm 0.07 ) \cdot 10^{0} $ & $( 1.15 \pm 0.50 ) \cdot 10^{2}$ \\
0.62 & $ ( 2.26 \pm 0.09 ) \cdot 10^{0} $ & $( 5.97 \pm 2.68 ) \cdot 10^{1}$ \\
0.66 & $ ( 1.28 \pm 0.04 ) \cdot 10^{0} $ & $( 2.73 \pm 1.82 ) \cdot 10^{1}$ \\
0.70 & $ ( 5.04 \pm 0.20 ) \cdot 10^{-1} $ & $( 9.87 \pm 8.67 ) \cdot 10^{0}$ \\
0.74 & $ ( 7.20 \pm 1.16 ) \cdot 10^{-2} $ & $( -0.70 \pm 3.08 ) \cdot 10^{0}$ \\
0.78 & $ ( 2.07 \pm 2.68 ) \cdot 10^{-4} $ & $( -2.04 \pm 2.80 ) \cdot 10^{-2}$ \\
0.82 & $ 0.00 $ & $ 0.00 $ \\
0.86 & $ 0.00 $ & $ 0.00 $ \\
0.90 & $ 0.00 $ & $ 0.00 $ \\
0.94 & $ 0.00 $ & $ 0.00 $ \\
0.98 & $ 0.00 $ & $ 0.00 $ \\ \hline
\end{tabular}
\end{center}
\caption{The Born level and next-to-leading order functions $B_{E_{3 jet}}$ and $C_{E_{3 jet}}$ for the softest-jet explanarity.}
\end{table}
The values for the functions $B_{E_{3 jet}}$ and $C_{E_{3 jet}}$ are given in table 4.
Figure 3 shows the distribution of the softest-jet explanarity.

\section{Conclusions}

In this paper we reported on the general purpose numerical 
program ``\mercutio'', which can
be used to calculate any infrared safe four-jet quantity in electron-positron
annihilation at next-to-leading order.
The program is based on the dipole formalism and uses a remapping of phase-space in order
to improve the efficiency of the Monte Carlo integration.
We presented 
numerical results for the four-jet fraction and the $D$-parameter.
These results agree with the results from other groups.
The program can also be used to investigate the internal structure of three-jet 
events at NLO. 
We also presented
 results for previously uncalculated observables: the jet broadening variable
and the softest-jet explanarity.

\end{document}